\documentstyle[aps]{revtex}
\begin{document}
\title{CP--Violation in $K$ and $B$ decays\thanks{%
This is an extended version of the talk given on 4th February 2003 at Sharif
University, Tehran, Iran, on the occasion of 16th Khwarizmi International
Award. I wish to thank Iranian Research Organization for Science and
Technology (IROST) for hospitality.}}
\author{Fayyazuddin}
\address{National Center for Physics\\
Quaid-e-Azam University Campus\\
Islamabad 45320 Pakistan}
\date{April 2003}
\maketitle

\begin{abstract}
In this article we give an overview of CP--violation for both $K^0\left( 
\bar{K}^0\right) $, $B^0\left( \bar{B}^0\right) $, and $B_s\left( \bar{B}%
_s\right) $ systems. Direct CP--violation and mixing induced CP--violation
are discussed for $K^0\left( \bar{K}^0\right) $, and $B^0\left( \bar{B}%
^0\right) $ decays.

(Report No.: NCP-QAU/0304-12)
\end{abstract}

\section{Introduction}

Symmetries have played an important role in particle physics. In quantum
mechanics a symmetry is associated with a group of transformations under
which a Lagrangian remains invariant. Symmetries limit the possible terms in
a Lagrangian and are associated with conservation laws. Here we will be
concerned with the role of discrete symmetries associated with space
reflection ($P$): $\vec{x}\rightarrow -\vec{x}$, Time reversal ($T$): $%
t\rightarrow -t$ and charge conjugation ($C$): transformation of particle to
its antiparticle.

Quantum Electrodynamics (QED) and Quantum Chromodynamics (QCD) respect all
these symmetries. Local quantum field theories in which Lorentz invariance
is built in are $CPT$ invariant. However in weak interactions $C$ and $P$
are maximally violated. $CP$ violation is a small effect observed in $B$ and 
$K$ decays. However the weak interaction Lagrangian in the standard model
violates $C$ and $P$ but is $CP$- invariant. Thus in the standard model the $%
CP$ non-conservation is a consequence of Higgs sector i.e. a mismatch
between the weak eigenstates and mass eigenstates in quark sector of the
standard model.

First we note that under $C$, $P$ and $T$ operations the Dirac spinor $\Psi $
transforms as follow 
\begin{eqnarray}
P\Psi \left( t,\vec{x}\right) P^{\dagger } &=&\gamma ^0\Psi \left( t,-\vec{x}%
\right)  \nonumber \\
C\Psi \left( t,\vec{x}\right) C^{\dagger } &=&-i\gamma ^2\gamma ^0\bar{\Psi}%
^T\left( t,\vec{x}\right)  \label{1} \\
T\Psi \left( t,\vec{x}\right) T^{\dagger } &=&\gamma ^1\gamma ^3\Psi \left(
-t,\vec{x}\right)  \nonumber
\end{eqnarray}
The effect of transformations $C$, $P$ and $CP$ on various quantities that
appear in a gauge theory Lagrangian are given below 
\[
\begin{array}{ccccc}
\text{Term/Transformation} & \text{Scalar} & \text{Pseudoscalar} & \text{%
Vector} & \text{Axial vector} \\ 
& \bar{\Psi}_i\Psi _j & i\bar{\Psi}_i\gamma _5\Psi _j & \bar{\Psi}_i\gamma
^\mu \Psi _j & \bar{\Psi}_i\gamma ^\mu \gamma ^5\Psi _j \\ 
P & \bar{\Psi}_i\Psi _j & -i\bar{\Psi}_i\gamma _5\Psi _j & \eta \left( \mu
\right) \bar{\Psi}_i\gamma ^\mu \Psi _j & -\eta \left( \mu \right) \bar{\Psi}%
_i\gamma ^\mu \gamma ^5\Psi _j \\ 
C & \bar{\Psi}_j\Psi _i & i\bar{\Psi}_j\gamma _5\Psi _i & -\bar{\Psi}%
_j\gamma ^\mu \Psi _i & \bar{\Psi}_j\gamma ^\mu \gamma ^5\Psi _i \\ 
CP & \bar{\Psi}_j\Psi _i & -i\bar{\Psi}_j\gamma _5\Psi _i & -\eta \left( \mu
\right) \bar{\Psi}_j\gamma ^\mu \Psi _i & -\eta \left( \mu \right) \bar{\Psi}%
_j\gamma ^\mu \gamma ^5\Psi _i
\end{array}
\]
The vector bosons associated with the electroweak unification group $%
SU_L\left( 2\right) \times U\left( 1\right) $ transform under $CP$ as 
\begin{eqnarray}
&&W_\mu ^{\pm }\left( \vec{x},t\right) \stackrel{CP}{\rightarrow }-\eta
\left( \mu \right) W_\mu ^{\mp }\left( -\vec{x},t\right)  \nonumber \\
&&Z_\mu \left( \vec{x},t\right) \stackrel{CP}{\rightarrow }-\eta \left( \mu
\right) Z_\mu \left( -\vec{x},t\right)  \label{2} \\
&&A_\mu \left( \vec{x},t\right) \stackrel{CP}{\rightarrow }-\eta \left( \mu
\right) A_\mu \left( -\vec{x},t\right)  \nonumber
\end{eqnarray}
\begin{eqnarray*}
\eta \left( \mu \right) &=&1\text{ \thinspace \thinspace \thinspace
\thinspace }\mu =0 \\
\eta \left( \mu \right) &=&-1\text{ \thinspace \thinspace \thinspace
\thinspace }\mu =1,2,3
\end{eqnarray*}
Then it is clear that electro-weak interaction Lagrangian is $CP$-invariant.

It is instructive to discuss the restrictions imposed by $CPT$ invariance. $%
CPT$ invariance implies 
\begin{eqnarray}
_{\text{out}}\left\langle f\left| {\cal L}\right| X\right\rangle &=&_{\text{%
out}}\left\langle f\left| \left( CPT\right) ^{-1}{\cal L}CPT\right|
X\right\rangle  \nonumber \\
&=&\eta _T^{x*}\eta _T^f\,\,_{\text{in}}\left\langle \tilde{f}\left| \left(
CP\right) ^{\dagger }{\cal L}^{\dagger }\left( CP\right) ^{-1\dagger
}\right| X\right\rangle ^{*}  \nonumber \\
&=&\eta _T^{x*}\eta _T^f\left\langle X\left| \left( CP\right) ^{-1}{\cal L}%
\left( CP\right) \right| f\right\rangle _{\text{in}}  \nonumber \\
&=&-\eta _T^{x*}\eta _T^f\eta _{CP}^f\left\langle \bar{X}\left| {\cal L}%
S_f\right| \bar{f}\right\rangle _{\text{out}}  \nonumber \\
&=&\eta _f\,\,_{\text{out}}\left\langle \bar{f}\left| S_f^{\dagger }{\cal L}%
^{\dagger }\right| \bar{X}\right\rangle ^{*}  \nonumber \\
&=&\eta _f\,\exp (2i\delta _f)_{\text{out}}\left\langle \bar{f}\left| {\cal L%
}\right| \bar{X}\right\rangle ^{*},\,\,\,\,{\cal L}^{\dagger }={\cal L}
\label{3}
\end{eqnarray}
Hence we get 
\begin{eqnarray}
_{\text{out}}\left\langle \bar{f}\left| {\cal L}\right| \bar{X}\right\rangle
&=&\eta _f\,\exp (2i\delta _f)_{\text{out}}\left\langle f\left| {\cal L}%
\right| X\right\rangle ^{*}  \nonumber \\
&=&\eta _f\exp (i\delta _f)A^{*}\left( X\rightarrow f\right)  \label{4}
\end{eqnarray}
In deriving the above result, we have put $\tilde{f}=f$ where $\symbol{126}$
means that momenta and spin are reversed. Since we are in the rest frame of $%
X$,$T$ will reverse only magnetic quantum number and we can drop $\symbol{126%
}$. Further we have used 
\begin{equation}
CP\left| X\right\rangle =-\left| X\right\rangle  \label{5}
\end{equation}
\begin{equation}
CP\left| f\right\rangle =\eta _f^{CP}\left| \bar{f}\right\rangle  \label{6}
\end{equation}
\begin{equation}
\left| f\right\rangle _{\text{out}}=S_f\left| f\right\rangle _{\text{in}%
}=\exp (2i\delta _f)\left| f\right\rangle _{\text{in}}  \label{7}
\end{equation}
$\delta _f$ is the strong interaction phase shift. If $CP$-invariance holds
then 
\[
_{\text{out}}\left\langle f\left| {\cal L}\right| X\right\rangle =_{\text{out%
}}\left\langle \bar{f}\left| {\cal L}\right| \bar{X}\right\rangle 
\]
i.e. 
\begin{equation}
A^{*}=A
\end{equation}
Thus necessary condition for $CP$-violation is that the decay amplitude $A$
should be complex. In view of our discussion above, schematically under $CP$
an operator $O\left( \vec{x},t\right) $ is replaced by 
\begin{equation}
O\left( \vec{x},t\right) \rightarrow O^{\dagger }\left( -\vec{x},t\right)
\label{9}
\end{equation}
The effective Lagrangian has the structure ($\,\,\,{\cal L}^{\dagger }={\cal %
L}$) 
\begin{equation}
{\cal L}=aO+a^{*}O^{\dagger }  \label{10}
\end{equation}
Hence $CP$-violation requires $a^{*}\neq a.$

Now $CP$-violation can also arise when the $CP$-eigenstates 
\begin{equation}
\left| X_{1,2}^0\right\rangle =\frac 1{\sqrt{2}}\left[ \left|
X^0\right\rangle \mp \left| \bar{X}^0\right\rangle \right]  \label{11}
\end{equation}
are not the mass eigenstates i.e. $CP$-violation in the mass matrix. $CP$-
violation due to mass mixing and in the decay amplitude has been
experimentally observed in $K^0$ and $B_d^0$. For $B_s$ decays, the $CP$%
-violation in the mass matrix is not expected in the Standard Model. In fact
time dependent $CP$-violation asymmetry gives a clear way to observe direct $%
CP$-violation in $B_s$ decays.

\section{${\bf K}^{0}{\bf -\bar{K}}^{0}{\bf \,\,}\,\,${\bf Complex}}

$K^{0}-\bar{K}^{0}$ complex in particle physics, is a simple example of two
states system in the sense that states $\left| K^{0}\right\rangle $ and $%
\left| \bar{K}^{0}\right\rangle $ form a complete set so that an arbitrary
state can expanded in terms of them. It has some interesting consequences
for $CP$-violation.

Weak Interactions: 
\begin{eqnarray*}
K^{0} &\rightarrow &\pi ^{+}\,\pi ^{-} \\
P\left( K^{0}\right) &=&-1,\,P\left( \pi ^{+}\,\pi ^{-}\right) =\left(
-1\right) ^{2}\left( -1\right) ^{l=0}=1
\end{eqnarray*}
Parity is not conserved.

Charge conjugation is also not conserved. 
\begin{eqnarray*}
&&\pi ^{+}\,\,\stackrel{C}{\rightarrow }\,\,\,\,\pi ^{-} \\
&&\mu ^{+}\,\,\stackrel{C}{\rightarrow }\,\mu ^{-}\,\, \\
&&\nu \stackrel{C}{\rightarrow }\bar{\nu}\,\,\,
\end{eqnarray*}
$\nu $ is left handed and $\bar{\nu}$ is right handed.

Helicity under $C$ and $P$ transforms as: 
\begin{eqnarray*}
{\cal H} &=&\frac{\vec{s}\cdot \vec{p}}{\left| \vec{p}\right| }\stackrel{C}{%
\rightarrow }{\cal H} \\
&&\stackrel{P}{\rightarrow }-{\cal H}
\end{eqnarray*}
Invariance under $C$ gives 
\[
\Gamma _{\pi ^{+}\rightarrow \mu ^{+}(-)\nu }=\Gamma _{\pi ^{-}\rightarrow
\mu ^{-}(-)\bar{\nu}}. 
\]

Experimentally 
\[
\Gamma _{\pi ^{+}\rightarrow \mu ^{+}(-)\nu }\gg \Gamma _{\pi
^{-}\rightarrow \mu ^{-}(-)\bar{\nu}} 
\]
showing that $C$ is violated in weak interactions. Under $CP$ 
\[
\Gamma _{\pi ^{+}\rightarrow \mu ^{+}(-)\nu }\stackrel{CP}{\rightarrow }%
\Gamma _{\pi ^{-}\rightarrow \mu ^{-}(+)\bar{\nu}} 
\]
which is seen experimentally.

Now 
\[
K^{0}\rightarrow \pi ^{+}\,\pi ^{-}\rightarrow \bar{K}^{0}\,\,,\,\,\left|
\Delta Y\right| =2 
\]
Thus weak interaction can mix $K^{0}$ and $\bar{K}^{0}$%
\[
\left\langle K^{0}\left| H\right| \bar{K}^{0}\right\rangle \neq 0. 
\]
Off diagonal matrix elements are not zero. Thus $K^{0}$ and $\bar{K}^{0}$
can not be mass eigenstates.

Select the phase: 
\[
CP\left| K^{0}\right\rangle =-\left| \bar{K}^{0}\right\rangle . 
\]

Define 
\begin{eqnarray*}
\left| K_1^0\right\rangle &=&\frac 1{\sqrt{2}}\left[ \left| K^0\right\rangle
-\left| \bar{K}^0\right\rangle \right] \\
\left| K_2^0\right\rangle &=&\frac 1{\sqrt{2}}\left[ \left| K^0\right\rangle
+\left| \bar{K}^0\right\rangle \right] \\
CP\left| K_1^0\right\rangle &=&\left| K_1^0\right\rangle \\
CP\left| K_2^0\right\rangle &=&-\left| K_2^0\right\rangle
\end{eqnarray*}
$K_1^0$ and $K_2^0$ are eigenstates of $CP$ with eigenvalues $+1$ and $-1$.

If $CP$ is conserved 
\begin{eqnarray*}
\left\langle K_2^0\left| H\right| K_1^0\right\rangle &=&\left\langle
K_2^0\left| \left( CP\right) ^{-1}H\left( CP\right) \right|
K_1^0\right\rangle \\
&=&-\left\langle K_2^0\left| H\right| K_1^0\right\rangle
\end{eqnarray*}
Therefore 
\[
\left\langle K_2^0\left| H\right| K_1^0\right\rangle =0. 
\]
Thus $\left| K_1^0\right\rangle $ and $\left| K_2^0\right\rangle $ can be
mass eigenstates.

$K_{1}^{0}\rightarrow \pi ^{+}\,\pi ^{-}$: large phase space; decay
probability large; short lived. 
\[
K_{2}^{0}\rightarrow \pi ^{+}\,\pi ^{-}\pi ^{0}\text{ small phase space,
Long lived.} 
\]
$K_{1}^{0}$ and $K_{2}^{0}$ are mass eigenstates, they form a complete set. 
\begin{eqnarray}
\left| \psi \left( t\right) \right\rangle &=&a\left( t\right) \left|
K_{1}\right\rangle +b\left( t\right) \left| K_{2}\right\rangle  \nonumber \\
\hbar &=&c=1  \nonumber \\
i\frac{d\left| \psi \left( t\right) \right\rangle }{dt} &=&\left( 
\begin{array}{cc}
m_{1}-\frac{i}{2}\Gamma _{1} & 0 \\ 
0 & m_{2}-\frac{i}{2}\Gamma _{2}
\end{array}
\right) \left| \psi \left( t\right) \right\rangle .  \label{2.1}
\end{eqnarray}
The Solution is 
\begin{eqnarray*}
a\left( t\right) &=&a\left( 0\right) \exp \left( -im_{1}t-\frac{1}{2}\Gamma
_{1}t\right) \\
b\left( t\right) &=&b\left( 0\right) \exp \left( -im_{2}t-\frac{1}{2}\Gamma
_{2}t\right)
\end{eqnarray*}
Suppose we start with $\left| K^{0}\right\rangle $ initially 
\[
\left| \psi \left( 0\right) \right\rangle =\left| K^{0}\right\rangle , 
\]
Then we get 
\begin{eqnarray*}
\left| \psi \left( t\right) \right\rangle &=&\frac{1}{\sqrt{2}}\left[ \exp
\left( -im_{1}t-\frac{1}{2}\Gamma _{1}t\right) \left| K_{1}\right\rangle
\right. \\
&&+\left. \exp \left( -im_{2}t-\frac{1}{2}\Gamma _{2}t\right) \left|
K_{2}\right\rangle \right]
\end{eqnarray*}
or 
\begin{eqnarray}
\left| \psi \left( t\right) \right\rangle &=&\frac{1}{\sqrt{2}}\left\{
\left[ \exp \left( -im_{1}t-\frac{1}{2}\Gamma _{1}t\right) \right. \right. 
\nonumber \\
&&\left. +\exp \left( -im_{2}t-\frac{1}{2}\Gamma _{2}t\right) \right] \left|
K^{0}\right\rangle  \nonumber \\
&&-\left[ \exp \left( -im_{1}t-\frac{1}{2}\Gamma _{1}t\right) \right. 
\nonumber \\
&&\left. \left. -\exp \left( -im_{2}t-\frac{1}{2}\Gamma _{2}t\right) \right]
\left| \bar{K}^{0}\right\rangle \right\}  \label{2.2}
\end{eqnarray}
However in $K^{0}-\bar{K}^{0}$ basis the mass matrix is given by 
\begin{eqnarray}
M &=&m-\frac{i}{2}\Gamma  \nonumber \\
&=&\left( 
\begin{array}{cc}
m_{11}-\frac{i}{2}\Gamma _{11} & m_{12}-\frac{i}{2}\Gamma _{12} \\ 
m_{21}-\frac{i}{2}\Gamma _{21} & m_{22}-\frac{i}{2}\Gamma _{22}
\end{array}
\right) .  \label{2.3}
\end{eqnarray}
Hermiticity of matrices $m_{\alpha \alpha ^{\prime }}$ and $\Gamma _{\alpha
\alpha ^{\prime }}$ gives 
\begin{eqnarray}
\left( m\right) _{\alpha \alpha ^{\prime }} &=&\left( m^{\dagger }\right)
_{\alpha \alpha ^{\prime }}=\left( m^{*}\right) _{\alpha ^{\prime }\alpha
},\,\,\Gamma _{\alpha \alpha ^{\prime }}=\,\Gamma _{\alpha ^{\prime }\alpha
}^{*}  \nonumber \\
\alpha &=&\alpha ^{\prime }=1,2  \label{2.4} \\
m_{21} &=&m_{12\,}^{*}\text{, }\,\Gamma _{21}=\Gamma _{12\,}^{*}  \nonumber
\end{eqnarray}
$CTP$ invariance gives 
\begin{eqnarray}
\left\langle K^{0}\left| M\right| K^{0}\right\rangle &=&\left\langle \bar{K}%
^{0}\left| M\right| \bar{K}^{0}\right\rangle  \nonumber \\
m_{11} &=&m_{22\,\,,}\,\Gamma _{11}=\Gamma _{22}  \label{2.5} \\
\left\langle \bar{K}^{0}\left| M\right| K^{0}\right\rangle &=&\left\langle 
\bar{K}^{0}\left| M\right| K^{0}\right\rangle \,\,\,\,\,\,\,\text{Identity.}
\nonumber
\end{eqnarray}
Diagonalization of mass matrix $M$ in eq. (\ref{2.3}) gives 
\begin{eqnarray}
m_{11}-\frac{i}{2}\Gamma _{11}-pq &=&m_{1}-\frac{i}{2}\Gamma _{1}  \nonumber
\\
m_{11}-\frac{i}{2}\Gamma _{11}+pq &=&m_{2}-\frac{i}{2}\Gamma _{2}
\label{2.6}
\end{eqnarray}
where 
\begin{eqnarray}
p^{2} &=&m_{12}-\frac{i}{2}\Gamma _{12}  \nonumber \\
q^{2} &=&m_{12}^{*}-\frac{i}{2}\Gamma _{12}^{*}  \label{2.7}
\end{eqnarray}
Assuming $CP$ conservation 
\begin{eqnarray}
\left\langle \bar{K}^{0}\left| M\right| K^{0}\right\rangle &=&\left\langle
K^{0}\left| M\right| \bar{K}^{0}\right\rangle  \nonumber \\
m_{21} &=&m_{12\,}\text{, }\Gamma _{21}=\Gamma _{12\,}  \label{2.8}
\end{eqnarray}
$m_{12\,}$ and $\Gamma _{12\,}$ are real. Thus 
\begin{eqnarray}
pq &=&m_{12}-\frac{i}{2}\Gamma _{12}  \nonumber \\
m_{1} &=&m_{11}-m_{12},\,\,\Gamma _{1}=\Gamma _{11}-\Gamma _{12}  \nonumber
\\
m_{2} &=&m_{11}+m_{12},\,\,\Gamma _{2}=\Gamma _{11}+\Gamma _{12}  \nonumber
\\
\Delta m &=&m_{2}-m_{1}=2m_{12}, \\
\,\Delta \Gamma &=&\Gamma _{2}-\Gamma _{1}=2\Gamma _{12}  \label{2.10}
\end{eqnarray}
However it was found experimentally that $CP$ is not conserved in $K^{0}$
decay. We note 
\begin{eqnarray*}
CP\left( K_{1}^{0}\right) &=&1 \\
CP\left( \pi ^{+}\,\pi ^{-}\right) &=&\left( -1\right) ^{l}\left( -1\right)
^{l}=1
\end{eqnarray*}
Thus 
\[
K_{1}^{0}\longrightarrow \pi ^{+}\,\pi ^{-} 
\]
is allowed by $CP$ conservation.

Experimentally it was found that long lived $K^{0}$ also decay to $\pi
^{+}\,\pi ^{-}$ but with very small probability. Small $CP$ non conservation
can be taken into account by defining 
\begin{eqnarray}
\left| K_{S}\right\rangle &=&\left| K_{1}^{0}\right\rangle +\varepsilon
\left| K_{2}^{0}\right\rangle  \nonumber \\
\left| K_{L}\right\rangle &=&\left| K_{2}^{0}\right\rangle +\varepsilon
\left| K_{1}^{0}\right\rangle  \label{2.11}
\end{eqnarray}
where $\varepsilon $ is a small number. Thus $CP$ non conservation manifest
itself by the ratio: 
\begin{eqnarray}
\eta _{+-} &=&\frac{A\left( K_{L}\rightarrow \pi ^{+}\,\pi ^{-}\right) }{%
A\left( K_{S}\rightarrow \pi ^{+}\,\pi ^{-}\right) }=\varepsilon
\label{2.12} \\
\left| \eta _{+-}\right| &\simeq &\left( 2.286\pm 0.017\right) \times 10^{-3}
\nonumber
\end{eqnarray}
Now $CP$ non conservation implies 
\begin{equation}
m_{12}\neq m_{12}^{*},\,\,\,\Gamma _{12}\neq \Gamma _{12}^{*}.  \label{2.13}
\end{equation}
Since $CP$ violation is a small effect 
\begin{eqnarray}
\mathop{\rm Im}
m_{12} &\ll &%
\mathop{\rm Re}
m_{12}  \nonumber \\
\mathop{\rm Im}
\Gamma _{12} &\ll &%
\mathop{\rm Re}
\Gamma _{12}  \label{2.14}
\end{eqnarray}
Further if $CP$- violation arises from mass matrix then 
\begin{equation}
\Gamma _{12}=\Gamma _{12}^{*}  \label{2.15}
\end{equation}

Thus $CP$- violation can result by a small term $i%
\mathop{\rm Im}
m_{12}$ in the mass matrix given in Eq. (\ref{2.1}). 
\begin{equation}
M=\left( 
\begin{array}{cc}
m_1-\frac i2\Gamma _1 & i%
\mathop{\rm Im}
m_{12} \\ 
-i%
\mathop{\rm Im}
m_{12} & m_2-\frac i2\Gamma _2
\end{array}
\right) .  \label{2.16}
\end{equation}
Diagonalization gives 
\begin{equation}
\varepsilon =\frac{i%
\mathop{\rm Im}
m_{12}}{\left( m_2-m_1\right) -i\left( \Gamma _2-\Gamma _1\right) /2}.
\label{2.17}
\end{equation}
Then from Eq. (\ref{2.10}) up to first order, we get 
\begin{eqnarray}
\Delta m &=&m_2-m_1\rightarrow m_{K_L}-m_{K_S}  \nonumber \\
&=&2%
\mathop{\rm Re}
m_{12}  \nonumber \\
\Delta \Gamma &=&\Gamma _2-\Gamma _1=\Gamma _L-\Gamma _S=2\Gamma _{12}
\label{2.18}
\end{eqnarray}
Then Eq. (\ref{2.2}) is unchanged; replace 
\begin{eqnarray*}
m_1 &\rightarrow &m_S,\,m_2\rightarrow m_L\, \\
\Gamma _1 &\rightarrow &\Gamma _S,\,\,\Gamma _2\rightarrow \Gamma _L
\end{eqnarray*}
Now 
\begin{eqnarray}
\Delta m &=&m_L-m_S  \nonumber \\
\Delta \Gamma &=&\Gamma _L-\Gamma _S  \nonumber \\
\Gamma _S &=&\frac \hbar {\tau _S}=7.367\times 10^{-12}\text{ MeV},\,\,\, 
\nonumber \\
&&\left. \tau _S=\left( 0.8935\pm 0.0008\right) \times 10^{-10}\text{ S}%
\right.  \nonumber \\
\Gamma _L &=&\frac \hbar {\tau _L}=1.273\times 10^{-14}\text{ MeV},\,\, 
\nonumber \\
&&\left. \,\tau _L=\left( 5.17\pm 0.04\right) \times 10^{-8}\text{ S}\right.
\nonumber \\
\Delta \Gamma &\simeq &-\Gamma _S  \nonumber \\
m_L &=&m+\frac 12\Delta m  \nonumber \\
m_S &=&m-\frac 12\Delta m  \label{2.19}
\end{eqnarray}
Hence from Eq. (\ref{2}) 
\begin{equation}
\left| \psi \left( t\right) \right\rangle \approx e^{\frac{-i}2mt}\left\{ 
\begin{array}{c}
\left[ e^{\frac{-1}2\Gamma _St}e^{\frac i2\Delta mt}+e^{-\frac i2\Delta
mt}\right] \left| K^0\right\rangle \\ 
-\left[ e^{\frac{-1}2\Gamma _St}e^{\frac i2\Delta mt}-e^{-\frac i2\Delta
mt}\right] \left| \bar{K}^0\right\rangle
\end{array}
\right\}  \label{2.20}
\end{equation}
Therefore probability of finding $\bar{K}^0$ at time $t$ [we started with $%
K^0$] 
\begin{eqnarray}
P\left( K^0\rightarrow \bar{K}^0,t\right) &=&\left| \left\langle \bar{K}%
^0\left| {}\right. \psi \left( t\right) \right\rangle \right| ^2  \nonumber
\\
&=&\frac 14\left( 1+e^{-\Gamma _St}-2e^{-\frac 12\Gamma _St}\cos \left(
\Delta m\right) t\right)  \nonumber \\
&=&\frac 14\left( 1+e^{-t/\tau _S}-2e^{-\frac 12t/\tau _S}\cos \left( \Delta
m\right) t\right)  \label{2.21}
\end{eqnarray}
If kaons were stable $(\tau _S\rightarrow \infty )$, then 
\begin{equation}
P\left( K^0\rightarrow \bar{K}^0,t\right) =\frac 12\left[ 1-\cos \left(
\Delta m\right) t\right]  \label{2.22}
\end{equation}
which shows that a state produced as pure $Y=1$ state at $t=0$ continuously
oscillates between $Y=1$ and $Y=-1$ state with frequency $\omega =\frac{%
\Delta m}\hbar $ and period of oscillation 
\begin{equation}
\tau =\frac{2\pi }{\left( \Delta m/\hbar \right) }.  \label{2.23}
\end{equation}
Kaons, however decay and oscillations are damped.

By measuring the period of oscillation, $\Delta m$ can be determined. 
\begin{equation}
\Delta m=m_{L}-m_{S}=\left( 3.489\pm 0.008\right) \times 10^{-12}\text{ MeV.}
\label{2.24}
\end{equation}
Such a small number is measured, consequence of superposition principle in
quantum mechanics 
\begin{eqnarray*}
\pi ^{-}p &\rightarrow &K^{0}\Lambda ^{0} \\
&&\left. ^{|}\!\!\!\longrightarrow \bar{K}^{0}p\rightarrow \pi ^{+}\Lambda
^{0}\right.
\end{eqnarray*}
$\pi ^{+}$ can only be produced by $\bar{K}^{0}$ in the final state. This
would give the clear indication of oscillation.

Coming back to $CP$-violation 
\begin{eqnarray}
\varepsilon &=&\frac{i%
\mathop{\rm Im}
m_{12}}{\Delta m-i\Delta \Gamma /2}\,,\,\,\varepsilon =\left| \varepsilon
\right| e^{i\phi _\varepsilon }  \label{2.25} \\
\tan \phi _\varepsilon &=&-2\Delta m/\Delta \Gamma =\Delta m/\Gamma
_S-\Gamma _L  \nonumber \\
&\approx &\frac{2\times 0.474\Gamma _S}{0.998\Gamma _S}  \nonumber \\
&\rightarrow &\phi _\varepsilon =43.49\pm 0.08^0C.  \label{2.26}
\end{eqnarray}
So far we have considered $CP$-violation due to mixing in the mass matrix.
It is important to detect the $CP$-violation in the decay amplitude if any.
This is done by looking for a difference between $CP$-violation for the
final $\pi ^0\pi ^0$ state and that for $\pi ^{+}\pi ^{-}$.Now due to Bose
stastics, the two pions can be either in $I=0$ or $I=2$ states. Using
Clebsch-Gorden (CG) coefficient 
\begin{eqnarray}
A\left( K^0\rightarrow \pi ^{+}\pi ^{-}\right) &=&\frac 1{\sqrt{3}}\left[ 
\sqrt{2}A_0e^{i\delta _0}+A_2e^{i\delta _2}\right]  \nonumber \\
A\left( K^0\rightarrow \pi ^0\pi ^0\right) &=&\frac 1{\sqrt{3}}\left[
A_0e^{i\delta _0}-\sqrt{2}A_2e^{i\delta _2}\right]  \label{2.27}
\end{eqnarray}
Now $CPT$-invariance [viz Eq. (\ref{2.14})] gives 
\begin{eqnarray}
A\left( \bar{K}^0\rightarrow \pi ^{+}\pi ^{-}\right) &=&\frac 1{\sqrt{3}}%
\left[ \sqrt{2}A_0^{*}e^{i\delta _0}+A_2^{*}e^{i\delta _2}\right]  \nonumber
\\
A\left( \bar{K}^0\rightarrow \pi ^0\pi ^0\right) &=&\frac 1{\sqrt{3}}\left[
A_0^{*}e^{i\delta _0}-\sqrt{2}A_2^{*}e^{i\delta _2}\right]  \label{2.28}
\end{eqnarray}
The dominant decay amplitude is $A_0$ due to $\Delta I=1/2$ rule, $\left|
A_2/A_0\right| \simeq 1/22$. Using the Wu and Yang phase convention, we can
take $A_0$ to be real, then neglecting terms of order $\varepsilon \frac{%
\mathop{\rm Re}
A_2}{A_0}$ and $\varepsilon \frac{%
\mathop{\rm Im}
A_2}{A_0}$ ,we get 
\begin{eqnarray}
\eta _{+-} &\equiv &\left| \eta _{+-}\right| e^{i\phi _{+-}}\simeq
\varepsilon +\varepsilon ^{\prime }  \nonumber \\
\eta _{00} &\equiv &\left| \eta _{00}\right| e^{i\phi _{00}}\simeq
\varepsilon -2\varepsilon ^{\prime }  \label{2.29}
\end{eqnarray}
where 
\begin{equation}
\varepsilon ^{\prime }=\frac i{\sqrt{2}}e^{i\left( \delta _2-\delta
_0\right) }%
\mathop{\rm Im}
\frac{A_2}{A_0}  \label{2.30}
\end{equation}
Clearly $\varepsilon ^{\prime }$ measures the $CP$-violation in the decay
amplitude, since $CP$-invariance implies $A_2$ to be real.

After $35$ years of experiments at Fermilab and CERN, results have converged
on a definitive non-zero result for $\varepsilon ^{\prime }$%
\[
R=\left| \frac{\eta _{00}}{\eta _{+-}}\right| ^{2}=\left| \frac{\varepsilon
-2\varepsilon ^{\prime }}{\varepsilon +\varepsilon ^{\prime }}\right|
^{2}\,\,,\,\varepsilon ^{\prime }\ll \varepsilon 
\]
\begin{eqnarray}
&\simeq &\left| 1-\frac{3\varepsilon ^{\prime }}{\varepsilon }\right|
^{2}\simeq 1-6%
\mathop{\rm Re}
\left( \varepsilon ^{\prime }/\varepsilon \right)  \nonumber \\
\mathop{\rm Re}
\left( \varepsilon ^{\prime }/\varepsilon \right) &=&\frac{1-R}{6}
\label{2.31} \\
&=&\left( 1.8\pm 0.4\right) \times 10^{-3}.  \label{2.32}
\end{eqnarray}
This is an evidence that although $\varepsilon ^{\prime }$ is a very small,
but $CP$-violation does occur in the decay amplitude. Further we note from
Eq. (\ref{2.30}) 
\[
\phi _{\varepsilon ^{\prime }}=\delta _{2}-\delta _{0}+\frac{\pi }{2}\approx
48\pm 4^{0} 
\]
where numerical value is based on an analysis of $\pi \pi $ scattering.

\section{${\bf B}^{0}{\bf -\bar{B}}^{0}{\bf \,\,}\,\,${\bf Complex}}

For $B^{0}$ meson, one finds (see below) 
\begin{eqnarray}
m_{12} &=&\left| m_{12}\right| ^{2i\phi _{M}}  \nonumber \\
\Gamma _{12} &=&\left| \Gamma _{12}\right| e^{2i\phi _{M}}  \label{3.1} \\
\left| \Gamma _{12}\right| &\ll &\left| m_{12}\right|  \label{3.2} \\
p^{2} &=&e^{2i\phi _{M}}\left[ \left| m_{12}\right| -i\left| \Gamma
_{12}\right| \right] \simeq \left| m_{12}\right| e^{2i\phi _{M}}  \nonumber
\\
q^{2} &=&e^{-2i\phi _{M}}\left[ \left| m_{12}\right| -i\left| \Gamma
_{12}\right| \right] \simeq \left| m_{12}\right| e^{-2i\phi _{M}}
\label{3.3} \\
pq &=&\left| m_{12}\right|  \nonumber
\end{eqnarray}
Hence the mass eigenstates $B_{H}^{0}$ and $B_{L}^{0}$ can be written as: 
\begin{eqnarray}
\left| B_{H}^{0}\right\rangle &=&\frac{1}{\sqrt{2}}\left[ \left|
B^{0}\right\rangle -e^{2i\phi _{M}}\left| \bar{B}^{0}\right\rangle \right]
\label{3.4} \\
\left| B_{L}^{0}\right\rangle &=&\frac{1}{\sqrt{2}}\left[ \left| \bar{B}%
^{0}\right\rangle +e^{2i\phi _{M}}\left| B^{0}\right\rangle \right]
\label{3.5}
\end{eqnarray}
$CP$ violation occurs due to phase factor $e^{2i\phi _{M}}$ in mass matrix.

Hence one gets [from Eq. (\ref{2.2})],

\begin{eqnarray}
\left| B^{0}\left( t\right) \right\rangle &=&\frac{1}{\sqrt{2}}\left\{
\left[ \exp \left( -im_{1}t-\frac{1}{2}\Gamma _{1}t\right) \right. \right. 
\nonumber \\
&&\,\left. +\exp \left( -im_{2}t-\frac{1}{2}\Gamma _{2}t\right) \right]
\left| B^{0}\right\rangle  \nonumber \\
&&-e^{+2i\phi _{M}}\left[ \exp \left( -im_{1}t-\frac{1}{2}\Gamma
_{1}t\right) \right.  \nonumber \\
&&\left. \left. -\exp \left( -im_{2}t-\frac{1}{2}\Gamma _{2}t\right) \right]
\left| \bar{B}^{0}\right\rangle \right\}  \nonumber \\
&&  \label{3.6}
\end{eqnarray}

For $B$-decays 
\begin{eqnarray}
\Gamma _1 &=&\Gamma _2=\Gamma  \nonumber \\
\Delta m_B &=&m_2-m_1  \nonumber \\
m &=&\frac 12\left( m_1+m_2\right) .  \label{3.7}
\end{eqnarray}
Then from (\ref{3.6}), we get 
\begin{eqnarray}
\left| B^0\left( t\right) \right\rangle &=&e^{-imt}e^{-\frac 12\Gamma
t}\left\{ \cos \left( \frac{\Delta m}2t\right) \left| B^0\right\rangle
\right.  \nonumber \\
&&\left. +ie^{+2i\phi _M}\sin \left( \frac{\Delta m}2t\right) \left| \bar{B}%
^0\right\rangle \right\}  \label{3.8}
\end{eqnarray}
Similarly we get 
\begin{eqnarray}
\left| \bar{B}^0\left( t\right) \right\rangle &=&-e^{-imt}e^{-\frac 12\Gamma
t}\left\{ \cos \left( \frac{\Delta m}2t\right) \left| \bar{B}^0\right\rangle
\right.  \nonumber \\
&&\left. +ie^{-2i\phi _M}\sin \left( \frac{\Delta m}2t\right) \left|
B^0\right\rangle \right\}  \label{3.9}
\end{eqnarray}
>From Eq. (\ref{3.8}) and (\ref{3.9}), the decay amplitudes for 
\begin{eqnarray}
B^0\left( t\right) &\rightarrow &f\,\,\,\,\,\,\,\,\,\,\,\,\,A_f\left(
t\right) =\left\langle f\left| H_w\right| B^0\left( t\right) \right\rangle 
\nonumber \\
\bar{B}^0\left( t\right) &\rightarrow &\bar{f}\,\,\,\,\,\,\,\,\,\,\,\,\bar{A}%
_{\bar{f}}\left( t\right) =\left\langle \bar{f}\left| H_w\right| \bar{B}%
^0\left( t\right) \right\rangle  \label{3.10}
\end{eqnarray}
are given by 
\begin{eqnarray}
\,\,\,A_f\left( t\right) &=&e^{-imt}e^{-\frac 12\Gamma t}\left\{ \cos \left( 
\frac{\Delta m}2t\right) A_f\right.  \nonumber \\
&&\,\,\,\,\,\,\,\,\left. +ie^{+2i\phi _M}\sin \left( \frac{\Delta m}2%
t\right) \bar{A}_{\bar{f}}\right\}  \label{3.11} \\
\bar{A}_{\bar{f}}\left( t\right) &=&e^{-imt}e^{-\frac 12\Gamma t}\left\{
\cos \left( \frac{\Delta m}2t\right) \bar{A}_{\bar{f}}\right.  \nonumber \\
&&\left. +ie^{-2i\phi _M}\sin \left( \frac{\Delta m}2t\right) A_f\right\} .
\label{3.12}
\end{eqnarray}
Consider the decay for which 
\[
CP\left| f\right\rangle =\eta _f\left| f\right\rangle 
\]
For this case we get, from Eqs. (\ref{3.11}) and (\ref{3.12}), 
\begin{eqnarray}
{\cal A}_f\left( t\right) &=&\frac{\Gamma _f\left( t\right) -\bar{\Gamma}%
_f\left( t\right) }{\Gamma _f\left( t\right) +\bar{\Gamma}_f\left( t\right) }%
=\cos \left( \Delta mt\right) \left( \left| A_f\right| ^2-\left| \bar{A}%
_f\right| ^2\right)  \nonumber \\
&&+i\sin \left( \Delta mt\right) \left( e^{2i\phi _M}A_f^{*}\bar{A}%
_f-e^{-2i\phi _M}A_f\bar{A}_f^{*}\right) /\left| A_f\right| ^2-\left| \bar{A}%
_f\right| ^2  \label{3.13}
\end{eqnarray}
For the above kind of decays which proceed through a single diagram (for
example tree graph), $\bar{A}_f/A_f$ is given by 
\[
\frac{\bar{A}_f}{A_f}=\frac{e^{i\left( \phi +\delta _f\right) }}{e^{i\left(
-\phi +\delta _f\right) }}=e^{2i\phi } 
\]
where $\phi $ is the weak phase in the decay amplitude. Hence from Eq. (\ref
{3.13}), we obtain 
\begin{equation}
{\cal A}_f=-\sin \left( \Delta mt\right) \sin \left( 2\phi _M+2\phi \right)
\label{3.14}
\end{equation}
In particular for the decay 
\[
B^0\rightarrow J/\psi \,K_s,\,\,\,\,\phi =0 
\]
we obtain 
\begin{equation}
{\cal A}_{\psi K_s}\left( t\right) =-\sin \left( 2\phi _M\right) \sin \left(
\Delta mt\right)  \label{3.15}
\end{equation}
and 
\begin{eqnarray}
{\cal A}_{\psi K_s} &=&\frac{\int_0^\infty \left[ \Gamma _f\left( t\right) -%
\bar{\Gamma}_f\left( t\right) \right] dt}{\int_0^\infty \left[ \Gamma
_f\left( t\right) +\bar{\Gamma}_f\left( t\right) \right] dt}  \nonumber \\
{\cal A}_{\psi K_s} &=&-\sin \left( 2\phi _M\right) \,\,\frac{\left( \Delta
m/\Gamma \right) }{1+\left( \Delta m/\Gamma \right) ^2}  \label{3.16} \\
\text{Expt.} &:&\left( \frac{\Delta m}\Gamma \right) _{B_d^0}=0.775\pm 0.015
\label{3.17}
\end{eqnarray}

The following comments are in order. There are three generations of
elementary fermions 
\begin{eqnarray*}
&&\left( 
\begin{array}{c}
u \\ 
d^{\prime }
\end{array}
\right) ,\,\left( 
\begin{array}{c}
\nu _{e} \\ 
e
\end{array}
\right) 
\begin{array}{lll}
& m_{e}=0.511\text{ MeV} &  \\ 
& m_{u},m_{d}\sim 4-5\text{ MeV} & 
\end{array}
\\
&&\left( 
\begin{array}{c}
c \\ 
s^{\prime }
\end{array}
\right) ,\left( 
\begin{array}{c}
\nu _{\mu } \\ 
\mu
\end{array}
\right) 
\begin{array}{lll}
& m_{\mu }=105\text{ MeV} &  \\ 
& m_{c}\sim 1.4\text{ GeV} &  \\ 
&  & 
\end{array}
\\
&&\,\left( 
\begin{array}{c}
t \\ 
b^{\prime }
\end{array}
\right) ,\left( 
\begin{array}{c}
\nu _{\tau } \\ 
\tau
\end{array}
\right) 
\begin{array}{lll}
& m_{\tau }=1.777\text{ GeV} &  \\ 
& m_{b}\approx 4-4.5\text{ GeV} &  \\ 
& m_{t}=175\text{ GeV} & 
\end{array}
\end{eqnarray*}
The left--handed fermions are put into doublet representation of electroweak
unification group SU$_{L}\left( 2\right) \times $U$\left( 1\right) $ as
follows 
\[
\begin{array}{cccccc}
&  &  & \!\!\!\!\!\!\!\!SU_{L}\left( 2\right) & I_{3L} & Y \\ 
\left( 
\begin{array}{c}
\nu _{e} \\ 
e
\end{array}
\right) _{L} & \left( 
\begin{array}{c}
\nu _{\mu } \\ 
\mu
\end{array}
\right) _{L} & \left( 
\begin{array}{c}
\nu _{\tau } \\ 
\tau
\end{array}
\right) _{L} & 2 & \frac{1}{2},-\frac{1}{2} & -1 \\ 
\left( 
\begin{array}{c}
u \\ 
d^{\prime }
\end{array}
\right) _{L} & \left( 
\begin{array}{c}
c \\ 
s^{\prime }
\end{array}
\right) _{L} & \left( 
\begin{array}{c}
t \\ 
b^{\prime }
\end{array}
\right) _{L} & 2 & \frac{1}{2},-\frac{1}{2} & \frac{1}{3} \\ 
e_{R} & \mu _{R} & \tau _{R} & 1 & 0 & -2 \\ 
u_{R} & c_{R} & t_{R} & 1 & 0 & \frac{4}{3} \\ 
d_{R} & s_{R} & b_{R} & 1 & 0 & -\frac{2}{3}
\end{array}
\]
\[
Q=I_{3L}+\frac{Y}{2} 
\]

Mediators of the weak interactions are put in the: adjoint representation of
SU$_{L}\left( 2\right) $%
\[
\begin{array}{cccc}
W_{\mu }^{+},W_{\mu }^{-},W_{\mu }^{0}, & : & \text{SU}_{L}\left( 2\right) & 
\text{Isospin} \\ 
B_{\mu } &  & \text{U}\left( 1\right) & \text{Hypercharge}
\end{array}
\]
Since weak forces are short range, the gauge group SU$_{L}\left( 2\right)
\times $U$\left( 1\right) $ is spontaneously broken: 
\[
\text{SU}_{L}\left( 2\right) \times \text{U}\left( 1\right) \longrightarrow 
\text{U}_{em}\left( 1\right) 
\]
The mass eigenstates are 
\begin{eqnarray*}
W^{+},W^{-},Z_{\mu } &=&-\sin \theta _{W}B_{\mu }+\cos \theta _{W}W_{\mu
}^{0} \\
A_{\mu } &=&\cos \theta _{W}B_{\mu }+\sin \theta _{W}W_{\mu }^{0}
\end{eqnarray*}
$W^{+},W^{-},Z_{\mu }$ are mediators of weak interactions. $A_{\mu }$:
photon mediator of electromagnetic interaction.

\begin{eqnarray*}
m_{A} &=&0,\text{ }m_{W}^{2}=\frac{\pi \alpha }{\sqrt{2}G_{F}\sin ^{2}\theta
_{W}},\,\,\,\,\sin ^{2}\theta _{W}\approx 0.23 \\
m_{Z} &=&\frac{m_{W}}{\cos \theta _{W}}
\end{eqnarray*}
After radiative corrections $m_{W}=80.39$ GeV, $m_{Z}=91.18$ GeV in
remarkable agreement with the experimental values.

We note that the weak eigenstates $d^{\prime },s^{\prime }$ and $b^{\prime }$
are not the same as mass eigenstates $d,s$ and $b$. 
\begin{equation}
\left( 
\begin{array}{c}
d^{\prime } \\ 
s^{\prime } \\ 
b^{\prime }
\end{array}
\right) =V\left( 
\begin{array}{c}
d \\ 
s \\ 
b
\end{array}
\right)  \label{3.18}
\end{equation}

$V$ is called the CKM matrix. 
\[
V=\left( 
\begin{array}{ccc}
V_{ud} & V_{us} & V_{ub} \\ 
V_{cd} & V_{cs} & V_{cb} \\ 
V_{td} & V_{ts} & V_{tb}
\end{array}
\right) 
\]
\begin{equation}
\simeq \left( 
\begin{array}{ccc}
1-\frac{1}{2}\lambda ^{2} & \lambda & A\lambda ^{3}\left( \rho -i\eta \right)
\\ 
-\lambda & 1-\frac{1}{2}\lambda ^{2} & A\lambda ^{2} \\ 
A\lambda ^{3}\left( 1-\rho -i\eta \right) & -A\lambda ^{2} & 1
\end{array}
\right) +O\left( \lambda ^{4}\right) ,\,\left. \lambda =0.22\right.
\label{3.19}
\end{equation}
The unitarity of $V$ 
\[
VV^{\dagger }=1 
\]
gives 
\begin{equation}
V_{ud}^{*}V_{ub}+V_{cb}^{*}V_{cd}+V_{td}^{*}V_{tb}=0  \label{3.20}
\end{equation}
The second line in Eq. (\ref{3.19}) expresses $V$ interms of Wolfenstein
parameterization. The unitarity of $V$ can be graphically represented as
triangle shown in Fig. 1.

\begin{eqnarray}
V_{cb} &=&A\lambda ^{2}  \nonumber \\
V_{ub} &=&\left| V_{ub}\right| e^{-i\gamma }  \nonumber \\
V_{td} &=&\left| V_{td}\right| e^{-i\beta }  \nonumber \\
\eta &:&\text{ Source of }CP\text{-violation}  \label{3.21}
\end{eqnarray}
We note that 
\begin{eqnarray}
\phi _{M} &=&\beta  \label{3.22} \\
A_{\psi K_{s}} &=&-\sin 2\beta \frac{\left( \Delta m/\Gamma \right) }{%
1+\left( \Delta m/\Gamma \right) ^{2}}  \label{3.23}
\end{eqnarray}
$A_{\psi K_{s}}$ has been experimentally measured which gives 
\[
\sin 2\beta =0.79\pm 0.14 
\]
But to see that $\phi _{M}=\beta $, we note that Fig. 2 represents the
transition $B^{0}\rightarrow \bar{B}^{0}$ ($t$--quark gives the leading
contribution) This transition is proportional to 
\begin{eqnarray}
\left( V_{tb}\right) ^{2}\left( V_{td}^{*}\right) ^{2} &=&\left|
V_{td}\right| ^{2}e^{2i\beta }  \nonumber \\
&=&A^{2}\lambda ^{6}\left[ \left( 1+\rho \right) ^{2}+\eta ^{2}\right]
e^{2i\beta }  \label{3.24} \\
m_{12} &=&\left| m_{12}\right| e^{2i\beta }  \nonumber \\
\Gamma _{12} &=&\left| \Gamma _{12}\right| e^{2i\beta }  \label{3.25}
\end{eqnarray}
Note that 
\[
\frac{\Gamma _{12}}{m_{12}}\sim \frac{m_{b}^{2}}{m_{t}^{2}} 
\]
hence $\Gamma _{12}\ll m_{12}$.

On the other hand: $B_s^0\rightarrow \bar{B}_s^0$ transition is proportional
to 
\begin{eqnarray}
\left( V_{tb}\right) ^2\left( V_{ts}^{*}\right) ^2 &=&\left| V_{ts}\right|
^2\approx A^2\lambda ^4  \label{3.26} \\
m_{12} &=&\left| m_{12}\right|  \nonumber \\
\Gamma _{12} &=&\left| \Gamma _{12}\right|  \label{3.27} \\
\phi _M &=&0  \label{3.28}
\end{eqnarray}
For $K$--mesons, 
\begin{eqnarray}
V_{ts}V_{td}^{*} &=&-A^2\lambda ^5\left( 1-\rho +i\eta \right)  \nonumber \\
V_{cs}V_{cd}^{*} &=&\left( 1-\frac 12\lambda ^2\right) \left[ -\lambda
+A^2\lambda ^5\left( 1-\rho +i\eta \right) \right]  \label{3.29}
\end{eqnarray}
It is clear from Eq. (\ref{3.29}) 
\[
\frac{\left( V_{ts}V_{td}^{*}\right) ^2}{\left( V_{cs}V_{cd}^{*}\right) ^2}=%
\frac{\lambda ^{10}m_t^2}{\lambda ^2m_c^2}\approx 5\times 10^{-6}\frac{m_t^2%
}{m_c^2} 
\]
Thus the relative contribution of $t$-quark as compared with the $c$ quark
is of the order 
\[
5\times 10^{-6}\frac{m_t^2}{m_c^2} 
\]
and hence is negligible for $m_t\approx 175$ GeV. Thus since $m_{12}$ and $%
\Gamma _{12}$ are proportional to the $\left( V_{cs}V_{cd}^{*}\right) ^2$ ,
we conclude from Eqs. (\ref{2.29}) that 
\begin{eqnarray*}
\mathop{\rm Im}
m_{12} &\ll &%
\mathop{\rm Re}
\left| m_{12}\right| \\
\mathop{\rm Im}
\Gamma _{12} &\ll &%
\mathop{\rm Re}
\left| \Gamma _{12}\right|
\end{eqnarray*}
These facts, we have used in discussing the $K^0-\bar{K}^0,B^0-\bar{B}^0$
systems. Still we need to determine $\gamma $, which directly measures the
CP--Violation in $B$ and $B_s$ decays.

\section{Direct $CP$-violation and final state interactions}

Diract $CP$-violation in $B$ decays involves the weak phase in the decay
amplitude. The reason for this being that necessary condition for direct $CP$%
-violation is that decay amplitude should be complex as discussed in section
1. But this is not sufficient because in the limit of no final state
interactions, the direct $CP$-violation in $B\rightarrow f$, $\bar{B}%
\rightarrow \bar{f}$ decay vanishes. To illustrate this point, we discuss
the decays $\bar{B}^{0}\rightarrow \pi ^{+}\pi ^{-}$. The main contribution
to this decay is from tree graph [Fig. 3]. But this decay can also proceed
via penguin diagram Fig. 4.

The contribution of penguin diagram can be written as 
\begin{equation}
P=V_{ub}V_{ud}^{*}f\left( u\right) +V_{cb}V_{cd}^{*}f\left( c\right)
+V_{tb}V_{td}^{*}f\left( t\right)  \label{4.1}
\end{equation}
where $f\left( u\right) $, $f\left( c\right) $ and $f\left( d\right) $
denote the contributions of $u$, $c$ and $t$ quarks in the loop. Now using
the unitarity equation (\ref{3.20}), we can rewrite Eq. (\ref{4.1}) as 
\begin{equation}
P=V_{ub}V_{ud}^{*}\left( f\left( u\right) -f\left( t\right) \right)
+V_{cb}V_{cd}^{*}\left( f\left( c\right) -f\left( t\right) \right)
\label{4.2}
\end{equation}
Due to loop integration $P$ is suppressed relative to $T$. But still its
contribution is not negligible. The first part of Eq. (\ref{4.2}) has the
same CKM matrix elements as for tree graph, so we can absorbe it in the tree
graph. Hence we can write $\left( f=\pi ^{+}\pi ^{-}\right) $%
\begin{equation}
\bar{A}_f=A\left( \bar{B}^0\rightarrow \pi ^{+}\pi ^{-}\right) =Te^{i\left(
\gamma +\delta _T\right) }+Pe^{i\left( \phi +\delta _P\right) }  \label{4.3}
\end{equation}
where $\delta _T$ and $\delta _P$ are strong interaction phases which have
been taken out so that $T$ and $P$ are real. $\phi $ is the weak phase in
Penguin graph; in fact it is zero for this particular decay. $CPT$
invariance gives 
\begin{equation}
A_f\equiv {}A\left( B^0\rightarrow \pi ^{+}\pi ^{-}\right) =Te^{-i\left(
\gamma -\delta _T\right) }+Pe^{-i\left( \phi -\delta _P\right) }.
\label{4.4}
\end{equation}
Hence direct $CP$-violation asymmetry is given by 
\begin{eqnarray}
C_{\pi \pi } &=&\frac{\Gamma \left( B^0\rightarrow \pi ^{+}\pi ^{-}\right)
-\Gamma \left( \bar{B}^0\rightarrow \pi ^{+}\pi ^{-}\right) }{\Gamma \left(
B^0\rightarrow \pi ^{+}\pi ^{-}\right) +\Gamma \left( \bar{B}^0\rightarrow
\pi ^{+}\pi ^{-}\right) }  \nonumber \\
&=&\frac{1-\left| \lambda \right| ^2}{1+\left| \lambda \right| ^2}  \nonumber
\\
&=&\frac{2r\sin \left( \delta _T-\delta _P\right) \sin \left( \phi -\gamma
\right) }{1+2r\cos \left( \delta _T-\delta _P\right) \cos \left( \gamma
-\phi \right) +r^2}  \nonumber \\
&=&\frac{2r\sin \left( \delta _T-\delta _P\right) \sin \gamma }{1+2r\cos
\left( \delta _T-\delta _P\right) \cos \gamma +r^2};\,\,r=\frac PT%
,\,\,\lambda =\frac{\bar{A}_f}{A_f}  \label{4.5}
\end{eqnarray}
We further note that (cf. Eqs. (\ref{3.4}) and (\ref{3.5})): 
\begin{eqnarray}
\frac{\Gamma \left( B_H^0\rightarrow \pi ^{+}\pi ^{-}\right) }{\Gamma \left(
B_L^0\rightarrow \pi ^{+}\pi ^{-}\right) } &\simeq &\tan ^2\left( \beta
+\gamma \right) \left[ 1+4r\frac{\sin \left( \phi -\gamma \right) }{\sin
\left( 2\beta +2\gamma \right) }\right] +O\left( r^2\right)  \nonumber \\
&=&\tan ^2\left( \beta +\gamma \right) \left[ 1-4r\frac{\sin \gamma }{\sin
\left( 2\beta +2\gamma \right) }\right]  \label{4.6}
\end{eqnarray}
This may be of academic interest; unless one can experimentally distinguish
between $B_H^0$ and $B_L^0$. For the time dependent $CP$-asymmetry for $%
B^0\rightarrow \pi ^{+}\pi ^{-}$ decay we obtain from Eqs. (\ref{3.13}) and (%
\ref{4.3}) 
\begin{mathletters}
\begin{equation}
{\cal A}(t)=C_{\pi \pi }(\cos \Delta mt)-S_{\pi \pi }(\sin \Delta mt),
\label{4.7a}
\end{equation}
where the direct CP--violation $C_{\pi \pi }$ is given in Eq. (4.5). The
mixing induced parameter $S_{\pi \pi }$ is given by 
\begin{equation}
S_{\pi \pi }=\frac{%
\mathop{\rm Im}
[e^{2i\phi _M}\lambda ]}{1+|\lambda |^2}=\frac{\sin \left( 2\beta +2\gamma
\right) +2r\cos \left( \delta _T-\delta _P\right) \sin \left( 2\beta +\gamma
\right) }{1+2r\cos \left( \delta _T-\delta _P\right) +r^2}  \label{4.7b}
\end{equation}
The recent BELLE and BABAR results are (see ref.) 
\end{mathletters}
\begin{eqnarray*}
\text{BELLE} &:&S_{\pi \pi }=-1.23\pm 0.41\text{ (stat) }\left. 
\begin{array}{c}
+0.08 \\ 
-0.07
\end{array}
\right. \text{ (syst.)} \\
\text{BABAR} &:&C_{\pi \pi }=-0.77\mp 0.27\text{ (stat) }\left. \mp
0.08\right. \text{ (syst.)}
\end{eqnarray*}
These results give clear indication of direct and mixing-induced $CP$%
-violation in $B^0\left( \bar{B}^0\right) \rightarrow \pi ^{+}\pi ^{-}$
decays. However it is clear from Eqs. (\ref{4.7a}) and (\ref{4.7b}), that
without the knowledge of strong interaction phases it is not possible to
extract weak interaction phases $\beta $ and $\gamma $ from the data.

In order to discuss the final state interactions (FSI) it is useful to
consider $\Delta C=\pm 1$, $\Delta S=-1$ decays of $B$. In $B$-decays, the $%
b $ quark is converted into $c$ or $u$-quark: 
\[
b\rightarrow c+q+\bar{q}\text{, }b\rightarrow u+q+\bar{q}. 
\]
For the $\Delta C=1$, $\Delta S=-1$ decays, $b\rightarrow c+q+\bar{q}$ is
relevant whereas for $\Delta C=-1$, $\Delta S=-1$ decays the transition $%
b\rightarrow u+q+\bar{q}$ enters.

In the tree graphs (See Fig. 5) the configuration is such that $q$ and $\bar{%
q}$ essentially go togather into the color singlet states, the third quark
recoiling, there is a significant probability that the system will hadronize
as a two body final state. In this picture final state interactions may be
neglected for the tree graphs (i.e. strong phase shifts are expected to be
small). The following decays proceed through tree graphs. They are dominant
decay modes and phase shifts for these decay modes are expected to be small. 
\begin{eqnarray}
&&\left. 
\begin{array}{lllll}
\bar{B}^{0}\rightarrow K^{-}D^{+}\,, &  & A_{-+}=a_{-+}e^{i\delta _{-+}} & :
& \,Te^{i\delta _{T}} \\ 
B^{-}\rightarrow K^{-}D^{0}, &  & A_{-0}=a_{-0}e^{i\delta _{-0}} & : & 
\,Te^{i\delta _{T}}+Ce^{i\delta _{C}} \\ 
\bar{B}_{s}^{0}\rightarrow K^{-}D_{s}^{+}\,, &  & B_{-s^{+}}=b_{s}e^{i\delta
_{s}} & : & Te^{i\delta _{T}}
\end{array}
\right] V_{cb}V_{us}^{*}  \label{4.7} \\
&&\left. 
\begin{array}{ccccc}
\bar{B}^{0}\rightarrow \pi ^{+}D_{s}^{-}\,, &  & \bar{A}_{-s^{+}}=\bar{a}%
e^{-i\delta \left( \gamma +\bar{\delta}\right) } & : & \bar{T}e^{i\left(
\delta _{T}+\gamma \right) } \\ 
\bar{B}_{s}^{0}\rightarrow K^{+}D_{s}^{-}\,, &  & B_{+s^{-}}=\bar{b}_{s}e^{i%
\bar{\delta}_{s}} & : & Te^{i\bar{\delta}_{T}}
\end{array}
\right] V_{ub}V_{cs}^{*}  \label{4.8}
\end{eqnarray}
\begin{eqnarray}
\frac{V_{ub}V_{cs}^{*}}{V_{cb}V_{us}^{*}} &=&\sqrt{\rho ^{2}+\eta ^{2}}%
e^{i\gamma },\qquad \sqrt{\rho ^{2}+\eta ^{2}}=0.36\pm 0.09  \label{4.9} \\
\frac{\bar{T}}{T} &\simeq &0.72\times \sqrt{\rho ^{2}+\eta ^{2}}
\label{4.10}
\end{eqnarray}

Color suppressed decays: (Fig. 6) are given below

\begin{eqnarray}
&& 
\begin{array}{lll}
\bar{B}^{0}\rightarrow \bar{K}^{0}D^{0}, &  & A_{00}=a_{00}e^{i\delta
_{00}}=\,\,Ce^{i\delta
_{c}}\,\,\,\,\,\,\,\,\,\,\,\,\,\,\,\,\,\,\,\,\,\,\,\,\,\,\,\,\,\,
\end{array}
V_{cb}V_{us}^{*}  \nonumber \\
&&\left. 
\begin{array}{lll}
\bar{B}^{0}\rightarrow \bar{K}^{0}\bar{D}^{0}, &  & \bar{A}%
_{00}=a_{00}e^{i\left( \bar{\delta}_{00}+\gamma \right) }=\,\,\bar{C}%
e^{i\left( \bar{\delta}_{c}+\gamma \right) } \\ 
B^{-}\rightarrow K^{-}\bar{D}^{0}, &  & \bar{A}_{-0}=a_{00}e^{i\left( \bar{%
\delta}_{00}+\gamma \right) }=\,\,\bar{C}e^{i\left( \bar{\delta}_{c}+\gamma
\right) }
\end{array}
\right] V_{ub}V_{cs}^{*}  \label{4.11}
\end{eqnarray}
\begin{equation}
\frac{C}{T}=\frac{\bar{C}}{\bar{T}}\simeq \left( \frac{a_{2}}{a_{1}}\right)
\approx 0.22  \label{4.12}
\end{equation}

We have neglected annihilation diagrams which are suppressed. For color
suppressed graphs the rescattering corrections may be important.
Rescattering is depicted in Fig.7

Rescattering is essentially determined by the scattering amplitudes of the
following processes. 
\begin{eqnarray*}
P_{a}+D &\rightarrow &P_{b}+D \\
P_{a}+\bar{D} &\rightarrow &P_{b}+\bar{D}
\end{eqnarray*}
$P$ stands for $K$ and $\pi $.

Using Regge phenomenology, one notes that pomeron, $\rho -A_2$ and $\omega
-f $ trajectories in $t$-channel contribute. Using $SU(3)$ and the fact that
these trajectories are exchange degenerate, one finds that the rescattering
corrctions can be expressed in terms of two parameters $\epsilon $ and $%
\theta $. For example, one finds (see Ref. [5])

\begin{eqnarray}
A\left( \bar{B}^{0}\rightarrow \bar{K}^{0}D^{0}\right) _{\text{FSI}}
&=&\epsilon e^{i\theta }A\left( \bar{B}^{0}\rightarrow K^{-}D^{+}\right) 
\nonumber \\
\epsilon &\simeq &0.08,\,\,\,\theta =73^{0}  \label{4.13}
\end{eqnarray}
After taking into account rescattering corrections, we get

\begin{mathletters}
\begin{eqnarray}
A_{00} &=&a_{00}e^{i\delta _{00}}+\epsilon e^{i\theta }a_{-+}e^{i\delta
_{-+}}  \nonumber \\
A_{-+} &=&a_{-+}e^{i\delta _{-+}}  \nonumber \\
A_{-0} &=&a_{-0}e^{i\delta _{-0}}+\epsilon e^{i\theta }a_{-+}e^{i\delta
_{-+}}  \label{4.14a} \\
B_{-s^{+}} &=&b_se^{i\delta _s}  \label{4.14b} \\
\bar{A}_{00} &=&\left( \bar{a}_{00}e^{i\bar{\delta}_{00}}+\epsilon
e^{i\theta }\bar{a}e^{i\bar{\delta}}\right) e^{i\gamma }  \nonumber \\
\bar{A}_{-0} &=&\left( \bar{a}_{-0}e^{i\bar{\delta}_{-0}}+\frac 12\epsilon
\left( 1-\frac i3\right) \bar{a}e^{i\bar{\delta}}\right) e^{i\gamma
}\,\,\,\,\,\,\,\,\,\,\,\,\,  \nonumber \\
\bar{A}_{0-} &=&\left( \bar{a}_{0-}e^{i\bar{\delta}_{0-}}+\frac 12\epsilon
\left( 1+\frac i3\right) \bar{a}e^{i\bar{\delta}}\right) e^{i\gamma }
\label{4.14c} \\
\bar{B}_{-s^{+}} &=&\bar{b}_se^{i\bar{\delta}_s}  \label{4.14d}
\end{eqnarray}
Further we obtain 
\end{mathletters}
\begin{equation}
\bar{\delta}_s=\delta _s\text{, \thinspace }\bar{\delta}=\delta _{-+},\,\,\,%
\bar{\delta}_{-0}=\bar{\delta}_{00}\text{,\thinspace \thinspace \thinspace
\thinspace \thinspace }\bar{\delta}_{00}=\delta _{00}  \label{4.15}
\end{equation}

In order to give some feeling, how the above results are obtained, we note
that using time reversal invariance, we get 
\begin{eqnarray}
A_f &\equiv &_{\text{out}}\left\langle f\left| H\right| B\right\rangle =_{%
\text{out}}\left\langle f\left| T^{-1}THT^{-1}T\right| B\right\rangle 
\nonumber \\
&=&_{\text{out}}\left\langle f\left| T^{-1}HT\right| B\right\rangle 
\nonumber \\
&=&_{\text{in}}\left\langle f^t\left| H^{\dagger }\right| B\right\rangle ^{*}
\nonumber \\
&=&_{\text{out}}\left\langle f\left| S^{\dagger }H\right| B\right\rangle
^{*}.  \label{4.16}
\end{eqnarray}
Hence 
\begin{eqnarray}
A_f^{*} &=&_{\text{out}}\left\langle f\left| S^{\dagger }H\right|
B\right\rangle =\sum_n\,_{\text{out}}\left\langle f\left| S^{\dagger
}\right| n\right\rangle _{\text{out\thinspace out}}\left\langle n\left|
H\right| B\right\rangle  \nonumber \\
&=&\sum_nS_{nf}^{*}\,A_n  \nonumber \\
&=&\sum_n\left( \delta _{nf}-2iM_{nf}^{*}\right) A_n  \label{4.17}
\end{eqnarray}
\begin{eqnarray}
A_f^{*}-A_f &=&-2i\sum_nM_{nf}^{*}A_n  \nonumber \\
\mathop{\rm Im}
A_f &=&\sum_nM_{nf}^{*}A_n  \label{4.18}
\end{eqnarray}
where $M_{nf}$ is the scattering amplitude for $f\rightarrow n$. The
dominant contribution to the decay amplitude in Eq. (\ref{4.18}) is from
those two body decays of $B$ which proceed through the channel $n\rightarrow
f^{\prime }$, where $A_{f^{\prime }}$ is given by tree graph (See Fig. 5).
In Eq. (\ref{4.18}), the two particle unitarity gives $%
\mathop{\rm Im}
A_f$ in terms of the scattering amplitude $M_{ff^{\prime }}$ and the tree
amplitude $A_{f^{\prime }}$ where the scattering amplitude $M_{ff^{\prime }}$
is obtained by using Regge phenomenology.. Then using unsubtracted
dispersion relation for the decay amplitude gives the rescattering
corrections to $A_f$ in the form $\varepsilon e^{i\theta }A_{f^{\prime }}$.

The relationship between the phase shifts given in Eq. (\ref{4.15}) follow
from the argument given below. First the equality $\bar{\delta}_s=\delta _s$%
, is the consequence of the $C-$invariance of strong interactions viz. 
\begin{eqnarray}
\left\langle K^{-}D_s^{+}\left| S\right| K^{-}D_s^{+}\right\rangle
&=&\left\langle K^{-}D_s^{+}\left| C^{-1}CSC^{-1}C\right|
K^{-}D_s^{+}\right\rangle  \nonumber \\
&=&\left\langle K^{-}D_s^{+}\left| C^{-1}SC\right| K^{-}D_s^{+}\right\rangle
\nonumber \\
&=&\left\langle K^{+}D_s^{-}\left| S\right| K^{+}D_s^{-}\right\rangle
\label{4.19}
\end{eqnarray}
For the rest of the relationships between phase shifts given in Eq. (\ref
{4.16}), we note that since $\gamma _{\rho D^{+}D^{-}\,}=-\gamma _{\omega
D^{+}D^{-}\,};$ the $\rho $ trajectory does not contribute to the scattering 
$K^{-}D^{+}\rightarrow K^{-}D^{+}$; $\rho $ trajectory also does not
contribute to the channel $\pi ^{+}D_s^{-}\rightarrow \pi ^{+}D_s^{-}$,
since $\rho -\omega $ are not compled to $D_s^{+}D_s^{-}$. Thus only pomeron
contribute to these channels and since pomeron is $SU\left( 3\right) $
singlet; hence it follows that $\bar{\delta}=\delta _{-+}$. Similarly since $%
-\gamma _{\rho K^0\bar{K}^0\,}=\gamma _{\omega K^0\bar{K}^0\,}$, only
pomeron contributes to the scattering channels $\bar{K}^0D^0\rightarrow \bar{%
K}^0D^0$, $\bar{K}^0\bar{D}^0\rightarrow \bar{K}^0\bar{D}^0$so that again $%
\delta _{00}=\bar{\delta}_{00}$. Similar argument holds for the equality of $%
\bar{\delta}_{-0}=\bar{\delta}_{0-}$

We now discuss the effect of FSI on $CP$-asymmetry. Define 
\[
D_{\mp }^0=\frac 12\left( D^0\mp \bar{D}^0\right) 
\]
$D_{\mp }^0$ are eigenstates of $CP$, with eigenvalue $\pm 1$. We define $CP$%
-asymmetry 
\begin{equation}
{\cal A}_{\mp }=\frac{\Gamma \left( B^{-}\rightarrow K^{-}D_{\mp }^0\right)
-\Gamma \left( B^{+}\rightarrow K^{+}D_{\mp }^0\right) }{\Gamma \left( \bar{B%
}^0\rightarrow K^{-}D^{+}\right) }  \label{4.20}
\end{equation}
Then one finds after taking out the weak interaction phase $e^{i\gamma }$: 
\begin{equation}
{\cal A}_{\mp }=\pm 2\sin \gamma \left[ \frac{%
\mathop{\rm Re}
A_{-0}%
\mathop{\rm Im}
\bar{A}_{-0}-%
\mathop{\rm Im}
A_{-0}%
\mathop{\rm Re}
\bar{A}_{-0}}{\left| A_{-+}\right| ^2}\right]  \label{4.21}
\end{equation}
Now using Eqs. (\ref{4.14a},\ref{4.14c} ), one gets 
\begin{eqnarray}
{\cal A}_{\mp } &=&\pm 2\sin \gamma \left[ -f\bar{r}\sin \left( \delta _{-0}-%
\bar{\delta}_{-0}\right) \right.  \nonumber \\
&&\;\;\;\;\;\;\;-\epsilon \bar{r}\sin \left( \theta +\delta _{-+}-\bar{\delta%
}_{-0}\right)  \nonumber \\
&&\,\,\,\,\,\,\,\,\,\,\,\,\,\,\,\,\,\,\,\,\,\,\,\left. +\frac{\sqrt{10}}6%
\epsilon f\bar{f}\sin \left( \theta -\phi +\bar{\delta}-\delta _{-0}\right)
\right] \text{, where}  \label{4.22} \\
f &=&\frac{a_{-0}}{a_{-+}}=\left( 1+\frac CT\right) \approx
1.22\,\,\,,\,\,\,\,\,\,\,\left( 1+\frac i{\sqrt{3}}\right) =\frac{\sqrt{10}}3%
e^{\pm \phi }  \nonumber \\
\bar{r} &=&\frac{\bar{a}_{-0}}{a_{-+}}=\left( \frac{\bar{C}}{\bar{T}}\right)
\left( \frac{\bar{T}}T\right) \,\,\,\,\,\,\,\phi =\tan ^{-1}1/3=18^0 
\nonumber \\
\bar{f} &=&\frac{\bar{a}}{a_{-+}}=\frac{\bar{T}}T  \label{4.23}
\end{eqnarray}
In the limit that the phase shifts $\delta $'s$\rightarrow 0$, the
rescattering corrections give 
\[
{\cal A}_{\mp }\sim 10^{-2}\sin \gamma 
\]
This may regarded as an upper limit. 
\begin{equation}
{\cal A}_{\mp }\leq 10^{-2}\sin \gamma  \label{4.24}
\end{equation}
Thus $A_{\mp }$ will be zero in the absence of FSI. A reliable estimate is
not possible, because it is not easy to estimate the strong interaction
phase shifts $\delta $'s. Our estimate for $\delta $'s gives 
\begin{equation}
{\cal A}_{\mp }\sim 10^{-3}\sin \gamma .
\end{equation}
For $B_s$ decays, defining 
\begin{equation}
B_{\mp }=\frac 1{\sqrt{2}}\left( B_s^0\mp \bar{B}_s^0\right) \text{,}
\label{4.25}
\end{equation}
we get 
\begin{eqnarray}
&&\frac{2\left| A\left( B_{\mp }\rightarrow K^{+}D_s^{-}\right) \right|
^2-b_s^2-\bar{b}_s^2}{2b_s\bar{b}_s}  \nonumber \\
&=&\mp \left[ \cos \left( \gamma -\delta _s+\bar{\delta}_s\right) \right] 
\nonumber \\
&=&\mp \left[ \cos \left( \gamma \right) \cos \left( \delta _s-\bar{\delta}%
_s\right) +\sin \left( \gamma \right) \sin \left( \delta _s-\bar{\delta}%
_s\right) \right]  \label{4.26}
\end{eqnarray}
\begin{eqnarray}
&&\frac{2\left| A\left( B_{\mp }\rightarrow K^{-}D_s^{+}\right) \right|
^2-b_s^2-\bar{b}_s^2}{2b_s\bar{b}_s}  \nonumber \\
&=&\mp \left[ \cos \left( \gamma +\delta _s-\bar{\delta}_s\right) \right] 
\nonumber \\
&=&\mp \left[ \cos \left( \gamma \right) \cos \left( \delta _s-\bar{\delta}%
_s\right) -\sin \left( \gamma \right) \sin \left( \delta _s-\bar{\delta}%
_s\right) \right]  \label{4.27}
\end{eqnarray}
Therefore 
\begin{eqnarray}
&&\Gamma \left( B_{\mp }\rightarrow K^{+}D_s^{-}\right) -\Gamma \left(
B_{\mp }\rightarrow K^{-}D_s^{+}\right)  \nonumber \\
&=&2b_s\bar{b}_s\sin \left( \gamma \right) \sin \left( \delta _s-\bar{\delta}%
_s\right)  \label{4.28}
\end{eqnarray}
Using $\delta _s=\bar{\delta}_s\,$, we get 
\begin{equation}
\,\,\Gamma \left( B_{\mp }\rightarrow K^{+}D_s^{-}\right) =\,\Gamma \left(
B_{\mp }\rightarrow K^{-}D_s^{+}\right)  \label{4.29}
\end{equation}

Finally we discuss the time dependent analysis of $B$-decays to get
information about weak phases, Define time--dependent $CP$-asymmetry
parameter: 
\begin{equation}
{\cal A}(t)\equiv \frac{[\Gamma _{f}(t)+\Gamma _{\bar{f}}(t)]-[\bar{\Gamma}%
_{f}(t)+\bar{\Gamma}_{\bar{f}}(t)]}{\Gamma _{f}(t)+\bar{\Gamma}_{f}(t)}
\label{4.30}
\end{equation}
For $f\equiv K_{s}D^{0}$ and $\bar{f}\equiv K_{s}\bar{D}^{0}$, using Eqs. (%
\ref{3.11} and \ref{3.12}) and Eqs. (\ref{4.14a},\ref{4.14c},\ref{4.15} ),
we get 
\begin{eqnarray}
{\cal A}\left( t\right) &=&-4\frac{\Gamma \left( \bar{B}^{0}\rightarrow
K^{-}D^{+}\right) (a_{2}/a_{1})(C/T)\sqrt{\rho ^{2}+\eta ^{2}}}{\Gamma
\left( \bar{B}^{0}\rightarrow K^{0}D^{0}\right) +\Gamma \left( \bar{B}%
^{0}\rightarrow K^{0}\bar{D}^{0}\right) }  \nonumber \\
&&\times \left[ \sin (\Delta m_{B}t)\sin (2\beta +\gamma )\right]  \nonumber
\\
&&\times \left[ 1+2\epsilon \left( a_{2}/a_{1}\right) \cos \theta +\epsilon
^{2}\left( a_{1}/a_{2}\right) \right]  \nonumber \\
&=&-4\frac{\sqrt{\rho ^{2}+\eta ^{2}}}{1+\rho ^{2}+\eta ^{2}}\left[ \sin
(\Delta m_{B}t)\sin (2\beta +\gamma )\right] \times 1.34  \nonumber \\
&=&-1.94\sin (\Delta m_{B}t)\sin (2\beta +\gamma )  \label{4.31}
\end{eqnarray}
Rescattering corrections are of the order of $34\%$.

For $B_{s}^{0}$, again using Eqs. (\ref{3.11} and \ref{3.12}), (\ref{4.14b},%
\ref{4.14d}), we get 
\begin{eqnarray}
{\cal A}_{s}\left( t\right) &\equiv &\frac{\Gamma _{fs}\left( t\right) -\bar{%
\Gamma}_{fs}\left( t\right) }{\Gamma _{fs}\left( t\right) +\bar{\Gamma}%
_{fs}\left( t\right) }  \nonumber \\
&=&\frac{b_{s}\bar{b}_{s}}{b_{s}^{2}+\bar{b}_{s}^{2}}\sin (\Delta
m_{B}t)\left[ S+\bar{S}\right]  \label{4.32} \\
f_{s} &\equiv &K^{+}D_{s}^{-}\text{,\thinspace \thinspace \thinspace }\bar{f}%
_{s}\equiv K^{-}D_{s}^{+}\text{\thinspace }  \nonumber \\
S &=&\sin \left( 2\phi _{\text{Ms}}+\gamma +\delta _{s}-\bar{\delta}%
_{s}\right)  \nonumber \\
\bar{S} &=&\sin \left( 2\phi _{\text{Ms}}+\gamma -\delta _{s}+\bar{\delta}%
_{s}\right)  \label{4.33}
\end{eqnarray}
Since for $B_{s}^{0}$, $\phi _{\text{Ms}}=0$ and $\delta _{s}=\bar{\delta}%
_{s}$, we get 
\begin{eqnarray}
{\cal A}_{s}\left( t\right) &\approx &\frac{2\sqrt{\rho ^{2}+\eta ^{2}}\bar{T%
}/T}{1+\left( \rho ^{2}+\eta ^{2}\right) \bar{T}/T}\sin (\Delta
m_{B_{s}}t)\sin \gamma  \nonumber \\
&\approx &0.49\sin \left( \Delta m_{B_{s}}t\right) \sin \gamma  \label{4.34}
\end{eqnarray}

We note that for time integrated CP-asymmetry 
\begin{eqnarray}
{\cal A}_s &\equiv &\frac{\int_0^\infty \left[ \Gamma _{fs}\left( t\right) -%
\bar{\Gamma}_{fs}\left( t\right) \right] dt}{\int_0^\infty \left[ \Gamma
_{fs}\left( t\right) +\bar{\Gamma}_{fs}\left( t\right) \right] dt}  \nonumber
\\
&=&\frac{b_s\bar{b}_s}{b_s^2+\bar{b}_s^2}\frac{\Delta m_{B_s}/\Gamma }{%
1+\left( \Delta m_{B_s}/\Gamma \right) ^2}\left[ S+\bar{S}\right] 
\label{4.35}
\end{eqnarray}
Thus for ${\cal A}_s$, we get 
\begin{equation}
{\cal A}_s\approx 0.49\,\,\sin \gamma \frac{\Delta m_{B_s}/\Gamma _s}{%
1+\left( \Delta m_{B_s}/\Gamma _s\right) ^2}  \label{4.36}
\end{equation}
The CP--asymmetry ${\cal A}_s\left( t\right) $ or ${\cal A}_s$ involves two
experimentally unknown parameters $\sin \gamma $ and $\Delta m_{B_s}$. Both
these parameters are of importance in order to test the unitarity of CKM
matrix viz whether CKM matrix is a sole source of CP--violation in the
processes in which CP--violation has been observed. We note that 
\begin{eqnarray}
\sin 2\beta  &=&\frac{2\eta \left( 1-\rho \right) }{\eta ^2+\left( 1-\rho
\right) ^2}  \label{4.37} \\
\sin \gamma  &=&\frac \eta {\sqrt{\eta ^2+\rho ^2}}  \label{4.38} \\
\frac{\Delta m_{B_s}}{\Delta m_{B_d}} &=&\frac{\left| V_{td}^2\right| }{%
\left| V_{ts}^2\right| }\xi =\frac 1{\lambda ^2\left[ \eta ^2+\left( 1-\rho
\right) ^2\right] }\xi   \label{4.39}
\end{eqnarray}
where $\xi $ is a measure of SU(3) violation; lattice calculation give its
value: 1.15$\pm 0.4.$

Experimental value for $\sin 2\beta $ is 0.79$\pm 0.14.$ Thus measurement of 
$\sin \gamma $ and $\Delta m_{B_{s}}$ will check the consistency of
CP-violation in B-decays.

\section{Conclusion}

Effective weak interaction Lagrangian in the standard model can accommodate $%
CP$-violation due to a mismatch between the weak eigenstates and mass
eigenstates These eigenstates are related by a unitary transformation given
by CKM matrix $V$. More than two generations of quarks are necessary to have
weak phases responsible for $CP$-violation. For three generations of quarks,
only two of phases $\alpha $, $\beta $ and $\gamma $ are independent because
unitarity of $V$ gives $\alpha +\beta +\gamma =\pi $. Both the direct and
mixing--induced CP--violation has been experimentally observed and has been
measured in $K$--decays. The mixing-induced CP--violation involves the mass
difference $\Delta m_K$, $\Delta m_B$ or $\Delta m_{B_s}${}and arises
because the mass eigenstates are not the CP--eigenstates. The mixing-induced
CP--violation in the decays $B^0\left( \bar{B}^0\right) \rightarrow
K_sJ/\psi ${} has been observed and $\sin \beta $ involving CKM phase $\beta 
$ has been measured. BELLE and BaBar groups have observed both direct and
mixing--induced CP--violation in $B^0\left( \bar{B}^0\right) \rightarrow \pi
^{+}\pi ^{-}$ decays and the parameters characterising these decays will be
experimentally determined more accurately shortly. However it must be
pointed out that direct CP--violation involves strong interaction phases due
to final state interactions. Thus it will not be easy to extract the weak
phase $\gamma $ in the decays $B^0\left( \bar{B}^0\right) \rightarrow \pi
^{+}\pi ^{-}$. In this respect the observation of mixing induced
CP--violation in the decay channels $\bar{B}_s^0\left( B_s^0\right)
\rightarrow K^{-}D_s^{+}\left( K^{+}D_s^{-}\right) $ and $\bar{B}_s^0\left(
B_s^0\right) \rightarrow K^{+}D_s^{-}\left( K^{-}D_s^{+}\right) $ which
involve the weak phase $\gamma $ and mixing parameter $\Delta m_{B_s}${}will
be of much interest. There is every indication that $CP$-violation in $K^0-%
\bar{K}^0$ and $B^0-\bar{B}^0$, $B_s^0-\bar{B}_s^0$ systems will be
understood in terms of CKM matrix. There is now strong evidence from
neutrino oscillations that in the lepton sector, weak eigenstates are also
related to mass eigenstates by a unitary transformation. Since for
neutrinos, it is experimentally established that there are three generations
of neutrinos, the unitary matrix connecting these states is a $3\times 3$
matrix. The form of this matrix is not fully known and therefore the
question of $CP$-violation in lepton sector is open one. Finally for baryon
genesis, both $C$ and $CP$-violation are required. How the $CP$-violation in
meson sector is related to $CP$-violation of baryogensis? There is no answer
to this question as yet.

It is a very selective list of refrences which have been consulted in the
prepration of this article

\section{Figure Captions}

\begin{enumerate}
\item  The CKM--Unitarity triangle

\item  Box diagrams for $\bar{B}^0\rightarrow B^0$ transition

\item  Tree graph for the decay $\bar{B}^0\rightarrow \pi ^{+}\pi ^{-}$

\item  Penguin graph for the decay $\bar{B}^0\rightarrow \pi ^{+}\pi ^{-}$

\item  Tree graph for $\Delta C=\pm 1$, $\Delta S=-1$ decays

\item  Color suppressed graphs for $\bar{B}^0\rightarrow \bar{K}^0D^0$, $%
\bar{B}^0\rightarrow \bar{K}^0\bar{D}^0$ decays

\item  Rescattering graphs for the decays $\bar{B}^0\rightarrow
K^{-}D^{+}\rightarrow \bar{K}^0D^0$, $\bar{B}^0\rightarrow \pi
^{+}D_s^{-}\rightarrow \bar{K}^0\bar{D}^0$ decays
\end{enumerate}

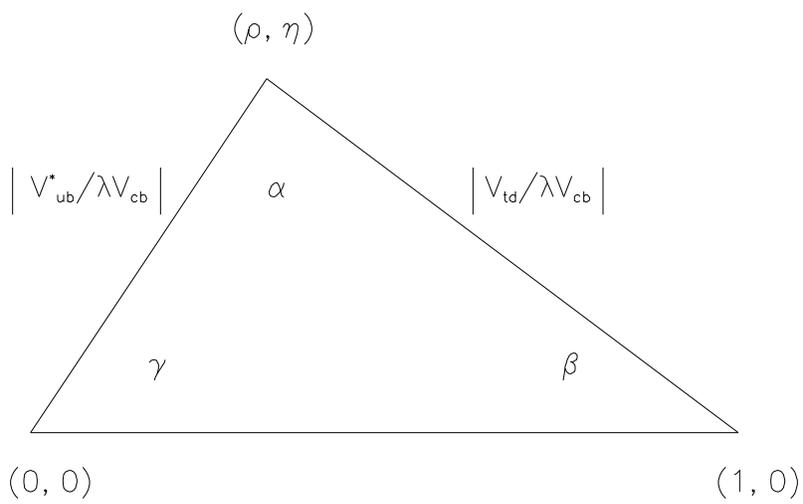
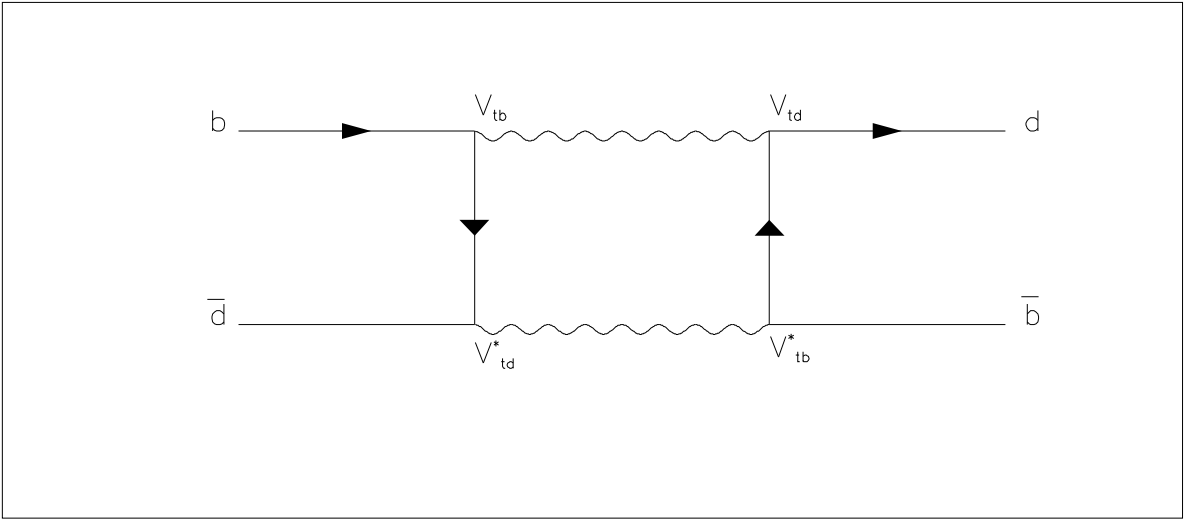
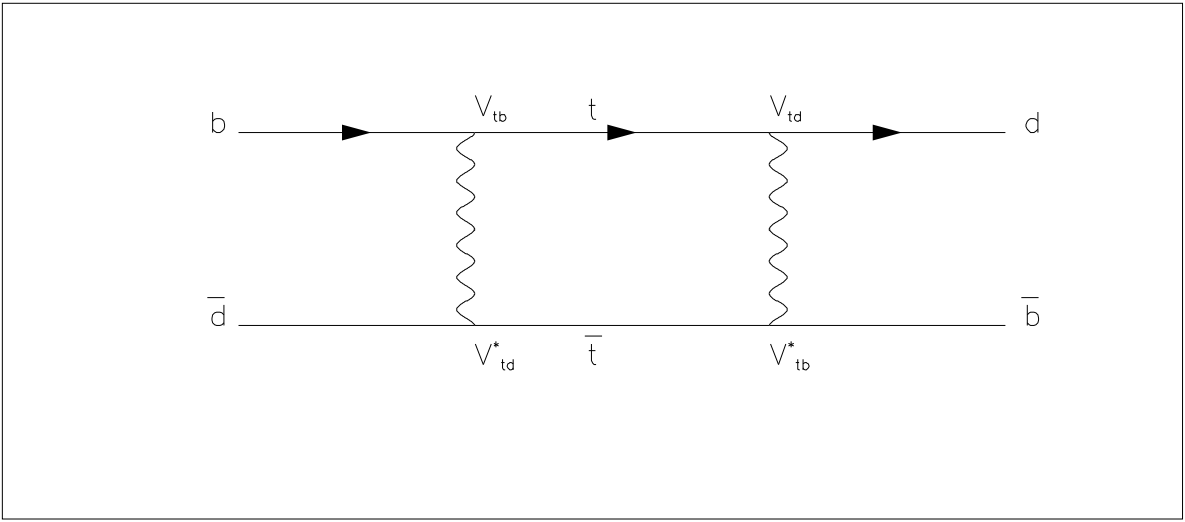
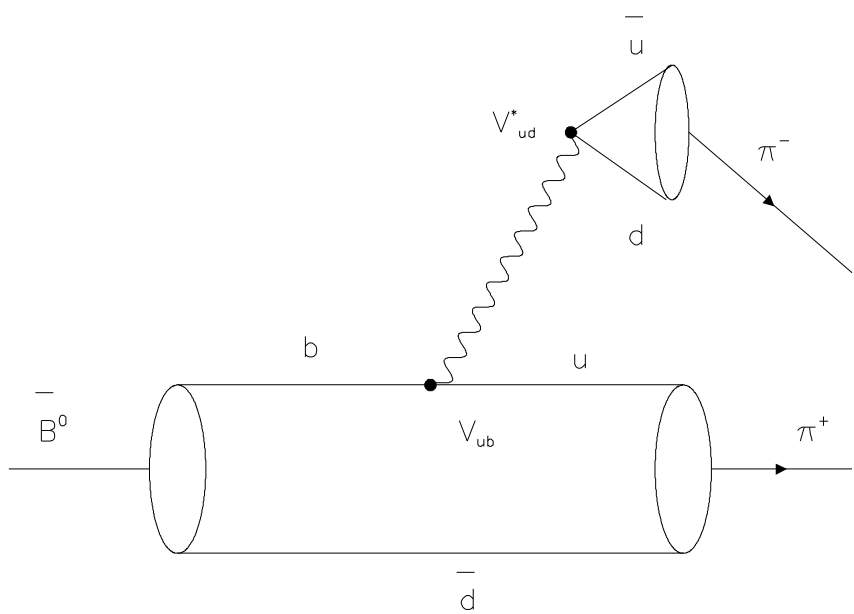
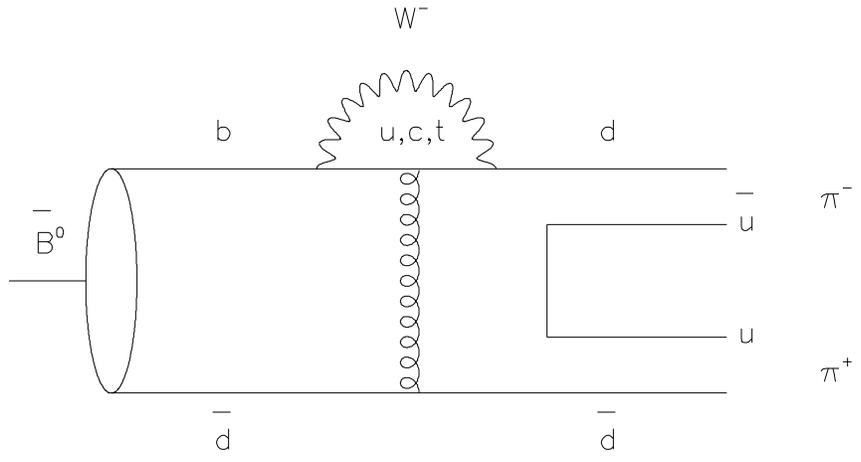
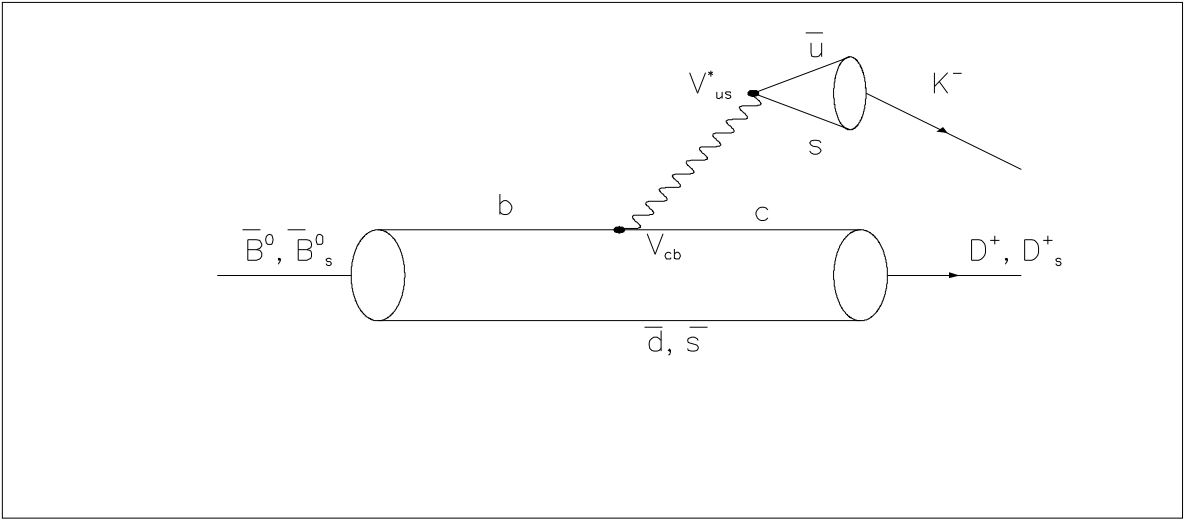
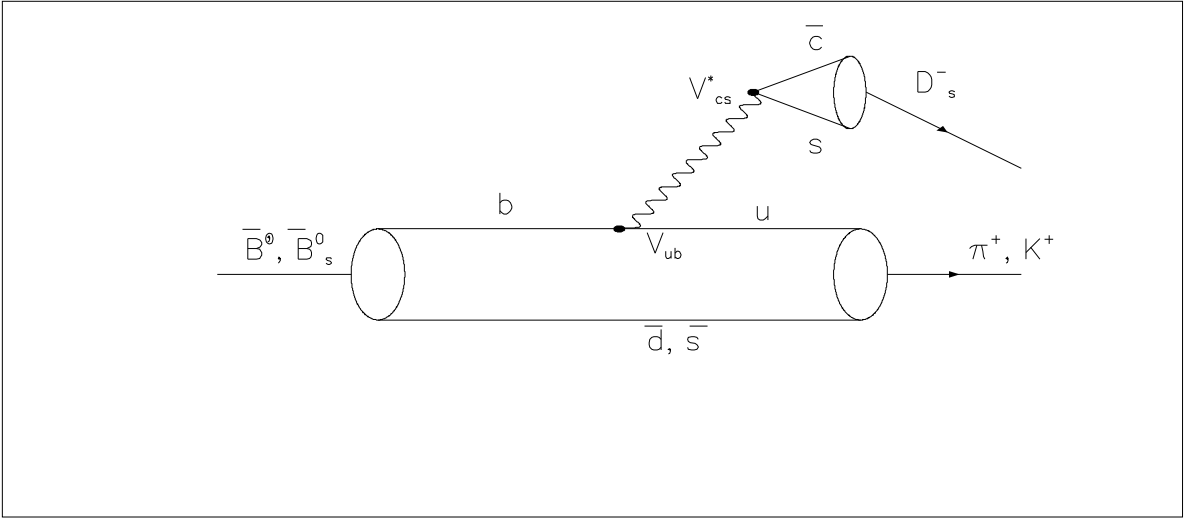
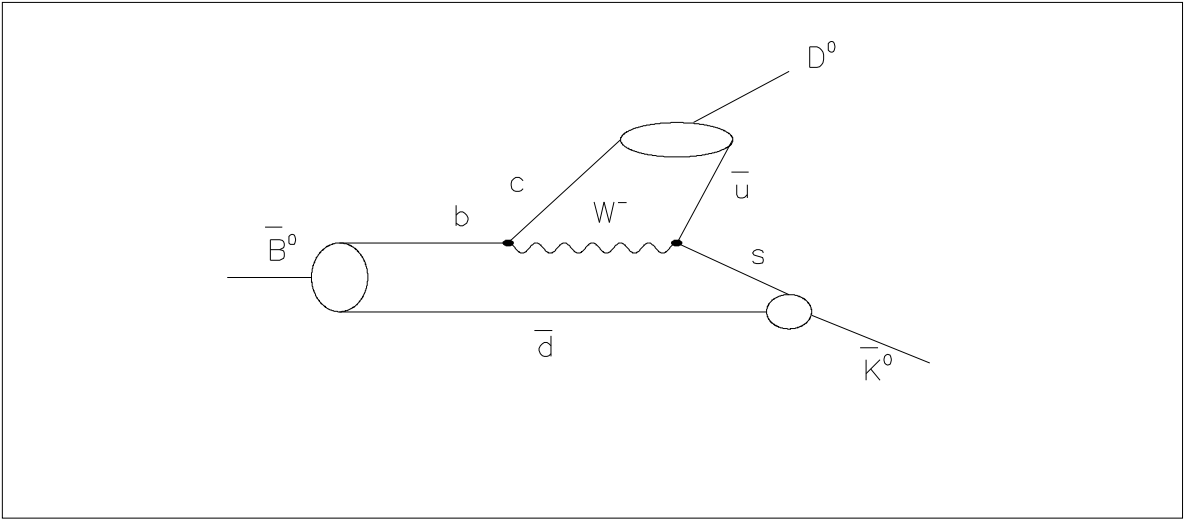
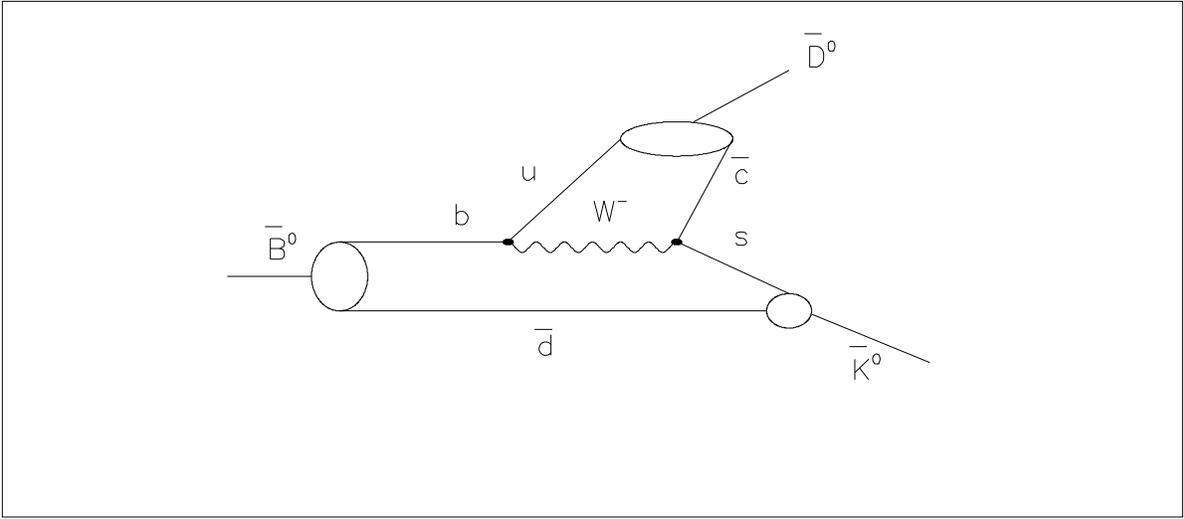
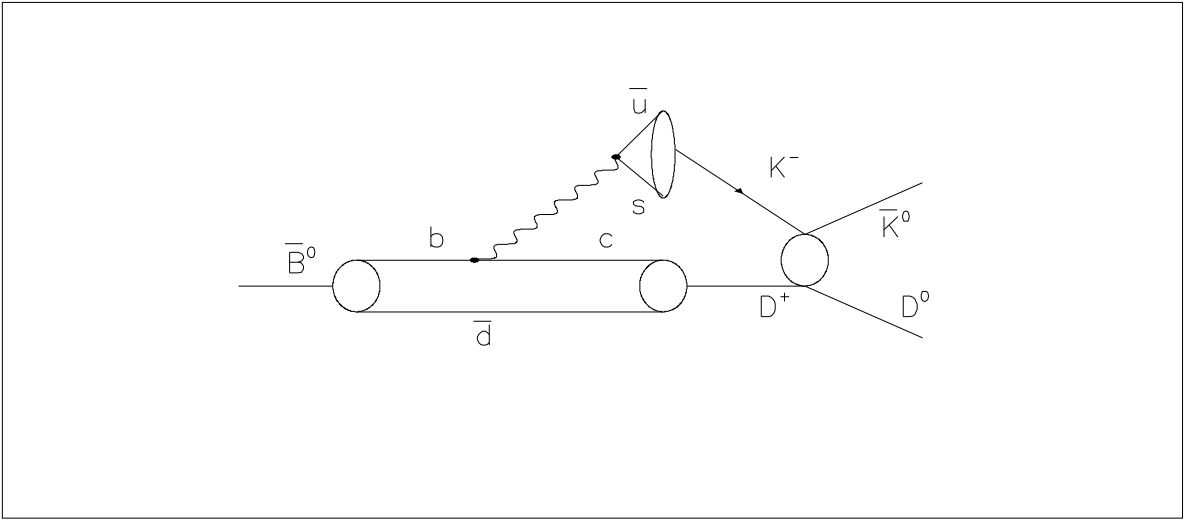
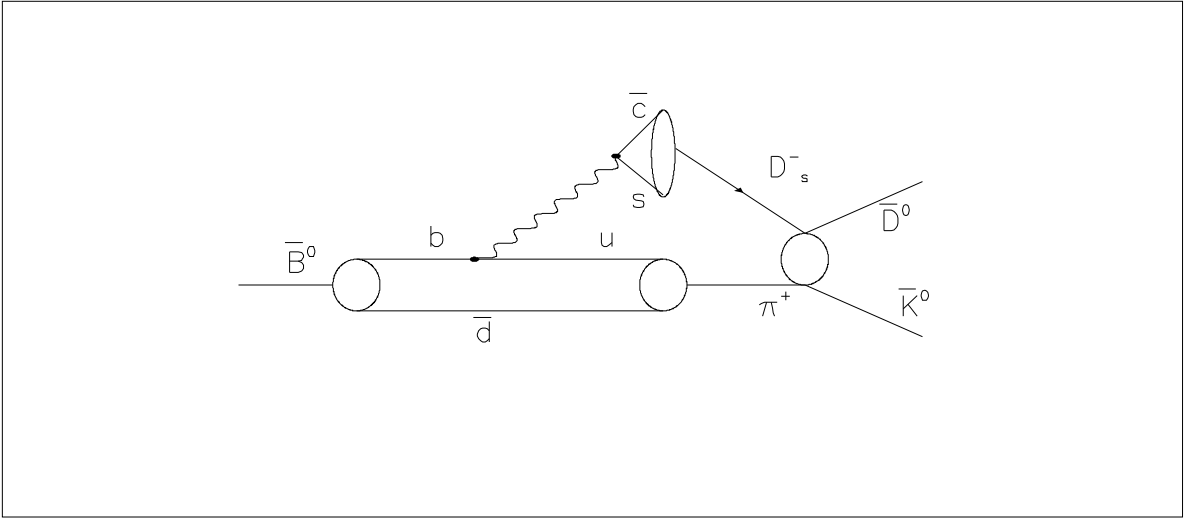
\end{document}